\documentclass[12pt]{article}
\usepackage{amsmath}
\usepackage{amssymb,amsfonts}

\catcode`@=11
\def\secteqno{\@addtoreset{equation}{section}\def\theequation{\thesection.\arabic{equation}}}
\catcode`@=12 \secteqno

\def\ebox#1#2{\vskip 2mm{\vbox{\hrule\hbox{\vrule\kern3pt\vbox{\kern3pt
         {\begin{eqnarray}#1\label{#2}\end{eqnarray}}
         \kern3pt}\kern3pt\vrule}\hrule}}\vskip 2mm}
\newcommand{\be}{\begin{equation}}
\newcommand{\ee}{\end{equation}}
\newcommand{\bea}{\begin{eqnarray}}
\newcommand{\eea}{\end{eqnarray}}

\def\6{\partial}
\def\7{\tilde}

\def\8{\widehat}

\def\={{\quad=\quad}}
\def\+{{\quad+\quad}}
\def\-{{\quad-\quad}}

\def\AdS5{{AdS$_5$}}
\def\S5{{ S$^5$ }}
\def\su224{{ SU(2,2$|$4)/(SO(4,2)$\times$SO(6)) }}

\def\G11{\Gamma_{11} }

\def\={{\;=\;}}
\def\+{{\;+\;}}

\def\-{{\;-\;}}

\begin{document}

\title{\hfill {\small CECS-PHY-06/04, UB-ECM-PF-06/07, TIT/HEP-556,
Toho-CP-0680}\\
$\mathcal{N}${\Large \textbf{=4 superconformal mechanics as a Non linear
Realization}}}
\author{\textbf{Andr\'{e}s Anabal\'{o}n$^{1,2}$}\thanks{%
anabalon-at-cecs-dot-cl}\textbf{\ \ \ \ Joaquim Gomis$^{1,3,4}$}\thanks{%
gomis-at-ecm-dot-ub-dot-es} \and \textbf{Kiyoshi Kamimura$^{5}$}\thanks{%
kamimura-at-ph-dot-sci-dot-toho-u-dot-ac-dot-jp} \ \ \ \textbf{Jorge Zanelli$%
^{1}$}\thanks{%
jz-at-cecs-dot-cl} \\
%EndAName
\textbf{\textit{$^{1}$}}{\small Centro de Estudios Cient\'{\i}ficos (CECS)
Casilla 1469 Valdivia, Chile}\\
{\small \textit{${}$}}\textbf{\textit{$^{2}$}}{\small Departmento de F\'{\i}%
sica, Universidad de Concepci\'{o}n Casilla 160-C, Concepci\'{o}n, Chile }$%
{} $\\
\textbf{\textit{$^{3}$}}{\small Departament ECM, Facultat de F\'{\i}sica,
Universitat de Barcelona, E-08028, Spain}\\
\textbf{\textit{$^{4}$}}{\small Department of Physics, Tokyo Institute of
Technology, Tokyo 152-8551, Japan}\\
\textbf{\textit{$^{5}$}}{\small Department of Physics, Toho University
Funabashi \ 274-8510, Japan}\\
}
\maketitle

\begin{abstract}
An action for a superconformal particle is constructed using the non linear
realization method for the group $PSU(1,1|2)$\textbf{,} without introducing
superfields. The connection between $PSU(1,1|2)$ and black hole physics is
discussed. The lagrangian contains six arbitrary constants and describes a
non-BPS superconformal particle. The BPS case is obtained if a precise
relation between the constants in the lagrangian is verified, which implies
that the action becomes kappa-symmetric.
\end{abstract}

\bigskip \thispagestyle{empty} \hfill

\vskip 15mm

\bigskip

\vskip10mm

\parskip=7pt

\section{Introduction}

The $\mathcal{N}=4$ superconformal symmetry appears in the dynamics of a
charged particle in the near horizon geometry of a four-dimensional charged
extremal black hole \cite{Claus:1998ts}. The connection can be traced back
to the geometry present in this case, which has the structure of $%
AdS_{2}\times S^{2}$. This implies that the mechanics describing the radial
motion of the charged particle in the near horizon geometry inherits the
global conformal symmetry group in one dimension, $SO(1,2)$, \cite%
{deAlfaro:1976je}. The near horizon geometry of the charged four-dimensional
extremal black hole is given by the Bertotti-Robinson metric\footnote{%
For a review see \cite{D'Auria:1998uh} \cite{Mohaupt:2000mj}},
\begin{equation}
ds^{2}=-\left( \frac{\rho }{M}\right) ^{2}dt^{2}+\left( \frac{M}{\rho }%
\right) ^{2}d\rho ^{2}+M^{2}d\Omega ^{2}.  \label{BRmetric}
\end{equation}%
This geometry admits two globally defined Killing spinors (it is a BPS state
with two local supersymmetries), which implies the existence of 8 real
supercharges. Hence, a simple mechanical model which captures this property
is the $PSU(1,1|2)$ conformal mechanics.\footnote{%
This group is often referred to as $SU(1,1|2)$, although this group contains
nontrivial central extensions, which are absent in $PSU(1,1|2)$.} The
superfield equations of motion can be constructed using the method of
nonlinear realizations (\textbf{NLR}) in superspace \cite{Ivanov:1988it}
\cite{deAzcarraga:1998ni}. $\mathcal{N}=4$ superconformal mechanics also
arises in the computation of the macroscopic black hole entropy of a D0-D4
black hole \cite{Gaiotto:2004ij}.

In this paper we will study more in general the dynamics of a superconformal
particle. This dynamical action is constructed by the method of non-linear
realizations without using superfields or requiring additional constraints
\cite{Gomis:2006xw}. As in \cite{Ivanov:1988it} \cite{deAzcarraga:1998ni},
we consider the coset $PSU(1,1|2)$ and no notion of geometry is used to
construct the action. The Goldstone fields will depend only on the world
line parameter $\tau $. This procedure allows us to consider in a unified
framework the cases of broken and unbroken supersymmetries. The lagrangian
will depend on six couplings constants, whose physical meaning is associated
with characteristics of the particle and the black hole, like mass, charge,
angular momentum. In the case with unbroken supersymmetries, a new local
gauge symmetry, kappa symmetry, appears so that half of the fermionic fields
can be gauged away and a BPS lagrangian is obtained. This symmetry appears
when the coupling constants verify a precise relation. This condition can be
understood in two ways, as an equality between the Casimir invariants of the
$SU(2)$ and the $SO(1,2)$ sectors, or more physically, as the equality $m=e$%
, where $m$ and $e$ are the mass and the charge of the particle. In this
case, the existence of supercharges, $Q$ and $S$, generating standard
supersymmetry and superconformal transformations, respectively, allows to
consider two kinds of BPS configurations; those that saturate the bound of
the hamiltonian $H=\left[ Q,Q^{\dagger }\right] _{+}$, and those saturating
the bound of the special conformal transformation generator $K=\left[
S,S^{\dagger }\right] _{+}$.

In the superconformal model considered here, there also appear two bosonic
local symmetries, one corresponds to ordinary diffeomorphisms of the world
line, and the other is a $U(1)$ gauge symmetry. The gauge symmetries
appearing in this model are understood as a right action of the coset
following reference \cite{Gomis:2006wu}. The $U(1)$ symmetry only transforms
the Goldstone fields associated to $SU(2)$ coordinates. Putting the fermions
to zero two decoupled lagrangians are obtained: i) the conformal mechanics
lagrangian written in a diffeomorphism invariant form, and ii) the
lagrangian of a particle on a sphere in which a monopole is located at the
center. The latter system has only one degree of freedom, in agreement with
the existence of the $U(1)$ gauge symmetry. If the fermions are switched on,
the two systems interact but the $U(1)$ symmetry still remains.

It is well known that, at the quantum level, the conformal mechanics has no
ground state associated to the hamiltonian $H$ and the wave function spreads
out to spatial infinity. In \cite{deAlfaro:1976je}, de Alfaro, Fubini and
Furlan suggested that one should consider the eigenstates of the compact
operator $P_{0}=\frac{1}{2}(H+K)$ which has a discrete spectrum and
normalizable eigenstates. From the perspective of the particle motion near
the black hole, this corresponds to a different choice of time \cite%
{Claus:1998ts}. In fact, the variable conjugate to $P_{0}$ is the global
time of AdS$_{2}$ and can describe the motion of the particle entering in
horizon, instead the time conjugate to $H$ only describes the motion of the
particle outside the horizon. Therefore, it is also natural to study the
dynamics of the superconformal particle using the new basis, that we call
the \emph{AdS basis}\footnote{%
A different parametrization for the AdS basis is used in \cite{Ivanov:2002tb}%
.}. In our approach this implies a new parametrization of the coset, leading
to a new parametrization for the action (see Section 5). The system is now
described by a relativistic lagrangian containing two square roots, plus a
WZ term that represents the coupling of the particle to the electromagnetic
field. This lagrangian has also three gauge invariances as in the previous
parametrization, also referred\textbf{\ }to as the \emph{conformal basis}.

In summary, the $\mathcal{N}=4$ super conformal model, which is presented
here in two different basis, describes the equatorial motion of a particle
in the background of a near horizon of a $\mathcal{N}=2$, charged,
four-dimensional, extremal black hole. A $D(2,1,\alpha)$ superconformal
mechanics in superfield formalism of \cite{Ivanov:2002pc} describes also a
motion of a particle in a equatorial plane. A general three-dimensional
motion in $\mathcal{N}=4$ conformal mechanics \cite{Gaiotto:2004ij}, \cite%
{Billo:1999ip}, \cite{Zhou:1999sm}\footnote{%
This author employs the non-linear realization approach with a different
coset, making use of the geometry of curves to construct the superconformal
action.}, \cite{Hatsuda:2003er} is not obtained with the coset considered
here. It is natural to ask whether there exits other cosets that can produce
a general three-dimensional motion without further physical or geometrical
requirements.

The outline of the paper is as follows. In section 2, the Maurer-Cartan (%
\textbf{MC}) forms are constructed, and in section 3 the Lagrangian in the
conformal basis is presented. In section 4 the gauge symmetries of the model
and the gauge fixed form of the lagrangian are studied, and in section 5 an
\emph{AdS} parametrization of the coset is given. Section 6 is devoted to
discussions. There are five appendices with some technical details.

\section{The $PSU(1,1|2)$ Lie algebra and its NLR}

The essential feature of the MC forms that make them useful objects to
describe dynamical systems is that they define invariants under a
non-linearly realized group action. The first step to calculate them is to
choose a coset, in this case, it follows from the discussion in the
introduction that the choice will be $PSU(1,1|2)$. The associated algebra is
formed by generators of dilatation $D$, special conformal transformations $K$%
, time translations $H$, $SU(2)$ rotations $J_{a}$, four supersymmetries $%
Q^{i}$, $Q_{i}^{\dagger }$, and four superconformal symmetries $S^{i}$, $%
S_{i}^{\dagger }$. The algebra is given in the appendix A.

Then it is possible to locally parametrize an arbitrary supergroup element $%
g $ as: \footnote{%
In the following, the index $i$ of the fermionic fields will not be written
explicitly.}
\begin{equation}
g=g_{0}e^{i(Q\eta ^{\dagger }+\eta Q^{\dagger })}e^{i(S\lambda ^{\dagger
}+\lambda S^{\dagger })}g_{J},\qquad g_{0}=e^{-itH}e^{izD}e^{i\omega
K},\quad g_{J}=e^{i\theta ^{1}J_{1}}e^{i\theta ^{2}J_{2}}e^{i\theta
^{3}J_{3}}.  \label{gpara}
\end{equation}%
In this approach, the coordinates $Z^{M}=\{t,z,\omega ,\eta ,\eta ^{\dagger
},\lambda ,\lambda ^{\dagger },\theta ^{a}\}$ in the group manifold will
become functions (Goldstone fields) of the worldline parameter $\tau $ --and
not superfields \cite{Gomis:2006xw}-- after the pullback on the world line
of the particle is taken. Note that we have also introduced a Goldstone
field, $t$, associated to the unbroken translation $H$. Here $g_{0}$ and $%
g_{J}$ parametrize the $SO(2,1)$ and the $SU(2)$ group elements,
respectively.

The left-invariant \textbf{(LI)} MC one-form $\Omega $ is given by
\begin{equation}
\Omega =-ig^{-1}dg=L^{H}H+L^{D}D+L^{K}K+QL^{\dagger Q}+L^{Q}Q^{\dagger
}+SL^{\dagger S}+L^{S}S^{\dagger }+L^{a}J_{a}=L^{A}G_{A},  \label{mcleft}
\end{equation}
where the one-forms $L^{A}=dZ^{M}L_{M}^{A}$ are given in appendix A. The MC
one-form $\Omega $ satisfies the MC equation
\begin{equation*}
d\Omega =-i\Omega \wedge \Omega ,
\end{equation*}
which merely asserts that (\ref{mcleft}) defines a flat connection. By
definition of the LI\ MC one forms $L^{A}$ are invariant under the left
action of the group. The explicit form of the infinitesimal group action on
the Goldstone fields is constructed in the next section.

\subsection{Global symmetry}

A mechanical system is defined by an action principle, which in this case is
given by the integral along the worldline of the pullback of the bosonic LI
MC forms. In order to characterize the states of the system, it is necessary
to identify the invariances of the action explicitly through the left
transformation of the Goldstone fields, $\delta _{L}Z^{M}$, under the
symmetry group.

As we have introduced all the generators to parametrize the group element $g$%
, each MC one-form component $L^{A}$ is invariant under global (rigid) group
transformations. The transformation of the Goldstone fields is defined from
the left action of the group on $g(Z^{M})$ as
\begin{equation}
g(Z^{M})\rightarrow e^{i\epsilon ^{A}G_{A}}g(Z^{M})=g(Z^{M}+\delta_L Z^{M}).
\end{equation}

At the level of the algebra, the left translations $\delta _{L}Z^{M}$, are
generated by the right-invariant (\textbf{RI}) vector fields $\widetilde{V}%
_{B}$, dual to the RI MC forms \cite{deAzcarraga:1995jw},
\begin{equation}
\tilde{\Omega}=-idgg^{-1}=dZ^{M}{{R_{M}}^{A}G_{A}}.
\end{equation}%
The RI vector fields $\widetilde{V}_{B}$ are related to the variations of
the Goldstone fields $\delta _{L}Z^{M}$ through
\begin{equation}
\widetilde{V}=\delta _{L}Z^{M}\frac{\partial }{\partial Z^{M}}=\epsilon
^{A}\;({R^{-1})_{A}{}^{M}}\frac{\partial }{\partial Z^{M}}=\epsilon ^{A}
\widetilde{V}_{A}.
\end{equation}%
This observation provides an alternative way to construct the $\delta
_{L}Z^{M}$. From the previous discussion it follows that the bosonic global
transformations for $PSU(1,1|2)$ are given by
\begin{eqnarray}
\text{Time translations} &:&\delta _{H}t=-\epsilon _{H},  \label{H} \\
\text{Dilatations} &\text{:}&\delta _{D}t=t\epsilon _{D},\quad \delta
_{D}z=\epsilon _{D},  \label{D} \\
\text{Special Conformal} &:&\delta _{K}t=-t^{2}\epsilon _{K},\quad \delta
_{K}z=-2t\epsilon _{K},\quad \delta _{K}\omega =e^{z}\epsilon _{K},
\label{K} \\
\text{SU(2) Rotations} &\text{:}&\delta _{SU(2)}\eta =-\frac{i}{2}\eta
\sigma _{a}\epsilon ^{a},\;\delta _{SU(2)}\lambda =-\frac{i}{2}\lambda
\sigma _{a}\epsilon ^{a},\; \\
&&\delta _{SU(2)}\theta ^{b}=(\mathbf{R}^{-1})_{\;a}^{b}\epsilon ^{a}.
\label{TD}
\end{eqnarray}%
The conjugate coordinates $\eta ^{\dagger }$ and $\lambda ^{\dagger }$
transform correspondingly. The matrix $(\mathbf{R}^{-1})_{\;a}^{b}$ is given
in appendix A and the supersymmetry transformations are in appendix B. In
the next section, the action principle is constructed.

\section{Dynamics of the superconformal mechanics}

The set of LI one-forms obtained from the Lie superalgebra $psu(1,1|2)$ can
be used as lagrangians for mechanical systems since they are, by definition,
objects that can be integrated along one dimensional trajectories. If we
assume an action with the lower number of derivatives,\footnote{%
Other combinations can be taken, for instance $\sqrt{b_{AB}L_{0}^{A}L_{0}^{B}%
}$, where $L^{A}=L_{0}^{A}d\tau $. In general this lagrangian will contains
accelerations eventually.} it is naturally given by a general linear
combination of the invariant one forms,
\begin{equation}
S=\int b_{A}\left( L^{A}\right) ^{\ast }d\tau,  \label{lagrangian}
\end{equation}%
where $\left( L^{A}\right) ^{\ast }$ stands for the pullback of $L^{A}$ to
the particle's worldline and the $b_{A}$'s are arbitrary coefficients.

It must be noted here that there is no a priori reason to rule out the
fermionic one-forms appropriately multiplied by Grassman numbers in order to
obtain the right Grassman parity for a bosonic action. For simplicity, this
possibility will not be considered here. The choice of only the bosonic LI
MC forms as lagrangians is the first physical assumption in the present
construction. Using (\ref{1})-(\ref{3}), the mechanical model invariant
under the $PSU(\left. 1,1\right\vert 2)$ group, constructed by taking the
pullback along a worldline parameter $\tau $, of a linear combination of the
bosonic one-forms $L^{H},L^{D},L^{K},L^{a}$ reads,
\begin{eqnarray}
\mathcal{S} &=&\int \mathcal{L}\;d\tau =\int \left(
b_{H}L^{H}+b_{D}L^{D}+b_{K}L^{K}+b_{a}L^{a}\right) ^{\ast }  \notag \\
&=&\int (L_{K}^{0})^{\ast }N_{K}+(L_{D}^{0})^{\ast }N_{D}+(L_{H}^{0})^{\ast
}N_{H}+b^{a}\left( L_{a}^{0}\right) ^{\ast }+N_{rest}^{\ast }\;.
\label{SUlagr}
\end{eqnarray}%
The coefficients $b_{A}$ are real but otherwise arbitrary, having the
dimensionalities $\left[ b_{H}\right] =l^{-1},\left[ b_{K}\right] =l^{1},%
\left[ b_{D}\right] =\left[ b_{a}\right] =l^{0}$. The $N_{H}$, $N_{D}$, $%
N_{K}$ and $N_{res}$ are defined in the appendix A by equations (\ref{NH}), (%
\ref{ND}), (\ref{NK}) and (\ref{Nres}), respectively. The $SU(2)$ coset
one-forms $L_{a}^{0}$ are given in (\ref{sucoset}), and the $SO(1,2)$ coset
forms $L_{K}^{0}$, $L_{D}^{0}$ and $L_{H}^{0}$ are given in (\ref{SOCOSET}).

By inspection of (\ref{SUlagr}) it can be noted that the velocity $\dot{%
\omega}$ appears, up to a boundary term, linearly in the lagrangian and
therefore $\omega $ can be eliminated from the action using its own equation
of motion,
\begin{equation}
\frac{\delta \mathcal{S}}{\delta \omega }=0\quad \Longrightarrow \quad
\omega =\frac{-\dot{N}_{K}-\dot{z}N_{K}+2e^{-z}\dot{t}N_{D}}{2e^{-z}\dot{t}%
N_{K}}\;.  \label{nondymanical}
\end{equation}%
Introducing the\textbf{\ }new bosonic coordinate $q$, defined by%
\begin{equation}
q=\sqrt{2}e^{z/2}\left( \frac{N_{K}}{b_{K}}\right) ^{1/2},  \label{kappaq}
\end{equation}%
the action (\ref{SUlagr}) can now be written as
\begin{equation}
\mathcal{S}=\int d\tau \;\left[ b_{K}\frac{\dot{q}^{2}}{2\dot{t}}-\frac{2%
\dot{t}}{b_{K}q^{2}}(N_{H}N_{K}-N_{D}^{2})-\frac{N_{D}\dot{N}_{K}}{N_{K}}%
\right] +N_{rest}^{\ast }+b^{a}(L_{a}^{0})^{\ast }.  \label{suinv}
\end{equation}

This action clearly resembles the conformal mechanics of \cite%
{deAlfaro:1976je}, with the characteristic $q^{-2}$ potential as the
interaction term which produces the nontrivial coupling between bosonic and
fermionic degrees of freedom.

One of the relevant aspects found in \cite{Claus:1998ts} is the explicit
relation between conformal mechanics and a physically nontrivial model
describing a charged particle in the near horizon geometry of an extremal,
four-dimensional Reissner-Nordstr\"{o}m black hole. Indeed, it is trivial to
show that the conformal mechanics of \cite{deAlfaro:1976je} \cite%
{Ivanov:1988vw} describes the motion of a particle on a background isometric
to $AdS_{2}$ . If the particle is charged, however, it would also interact
with the electromagnetic field of the black hole, and the trajectory would
no longer be a geodesic of the manifold.

In order to compare with Ref. \cite{Claus:1998ts}, it is enlightening to
write down the purely bosonic part of the action.
\begin{equation}
\left. \mathcal{S}\right\vert _{\eta =\eta ^{\dagger }=\lambda =\lambda
^{\dagger }=0}=\int d\tau \left[ b_{K}\frac{\dot{q}^{2}}{2\dot{t}}-\frac{2%
\dot{t}}{q^{2}}\left( \frac{b_{H}b_{K}-b_{D}^{2}}{b_{K}}\right) \right] +{%
b^{a}}(L_{a}^{0})^{\ast },  \label{blag}
\end{equation}%
which explicitly reflects the global invariance under the bosonic part of $%
PSU(1,1|2)$, namely, $SO(1,2)\times SU(2)$. As $\dot{\theta}^{3}$ enters
linearly in $b_{a}(L_{a}^{0})^{\ast }$, see (\ref{sucoset}), the $\theta
^{3} $ coordinate can be eliminated as well by using its own equation of
motion. The resulting action reads

\begin{eqnarray}
\left. \mathcal{S}\right\vert _{\eta =\eta ^{\dagger }=\lambda =\lambda
^{\dagger }=0} &=&\int d\tau \left[ b_{K}\frac{\dot{q}^{2}}{2\dot{t}}-\frac{2%
\dot{t}}{q^{2}}\left( \frac{b_{H}b_{K}-b_{D}^{2}}{b_{K}}\right) +\sqrt{{%
b_{1}^{2}+b_{2}^{2}}}\sqrt{\dot{\theta}_{1}^{2}\cos ^{2}\theta _{2}+\dot{%
\theta}_{2}^{2}}\right.  \notag \\
&&\left. -{b_{3}}\;\dot{\theta}_{1}\sin \theta _{2}\right] .  \label{blag1}
\\
&=&\int d\tau [L(q)+L(\theta ^{a})]  \label{blag2}
\end{eqnarray}
The direct product geometry of the BR metric (\ref{BRmetric}) is reflected
in the first three terms. They represent a geodesic in $AdS_{2}$ and a
geodesic in $S^{2}$. The last term can be interpreted, following \cite%
{Plyushchay:2000hb}, as the {electric} coupling of the particle with a
monopole field located at the center of $S^{2}$. Further physical life can
be given to this model, comparing $L(q)$ and $L(\theta ^{a})$ of (\ref{blag2}%
) with equation ($2.11$) of \cite{Claus:1998ts} and equation (8.4) of \cite%
{Plyushchay:2000hb} respectively, the constants $b_{A}$ can be identified as
\begin{equation}
b_{K}=m\qquad \left( b_{H}b_{K}-b_{D}^{2}\right) =2M^{2}\left( m-e\right)
m+J^{2}\qquad b^{a}b^{a}=J^{2}\qquad b_{3}=ge.  \label{coe}
\end{equation}
Here $m$ is the mass, $e$ the electric charge and $J$ is the angular
momentum of the particle, while $M$ is the black hole mass and $g$ is the
monopole charge. The authors of \cite{Claus:1998ts} used the constant
angular momentum on shell condition, replacing it in the equation of motion
of q and, in advance of quantization, wrote the angular momentum as $l(l+1)$%
. In the identification (\ref{coe}) this convention has not been followed.

The appearance of a monopole field has its roots in the fact that $SU(2)$ is
homeomorphic to $S^{3}$, since an atlas over $S^{3}$ defines a fiber bundle
(the Hopf bundle) classified by the transition function in the $n=1$
homotopy class of $\pi _{1}(U(1))=Z$. This is identical to the
characterization of a magnetic monopole of unit strength.

An interesting mechanism has operated here: the elimination of some non
dynamical variables from their equations of motion produced a recombination
of the $b^{A}$'s among themselves, giving rise to the effective parameters
of the theory (\ref{blag1}).

The relation between the parameters of the conformal mechanics and
observables have a nice example in the de Alfaro, Fubini and Furlan
conformal mechanics \cite{deAlfaro:1976je}, where the coefficient $g$ in the
hamiltonian
\begin{equation}
H=\frac{1}{2}\left( p^{2}+\frac{g}{q^{2}}\right)
\end{equation}%
can be recognized as the Casimir operator of the conformal group $SO(2,1)$,
classifying the irreducible representations of that group \cite%
{deAlfaro:1976je}.

In the action (\ref{blag1}) there is no direct coupling between the bosonic
coordinates $q$ and $\theta ^{a}$; they interact only through the fermions.

\section{Local Symmetries}

\subsection{Local symmetries in general}

\ In order to examine the local symmetries of the action (\ref{SUlagr})
using the NLR approach we followed the procedure developed in \cite%
{Gomis:2006wu}. The gauge symmetries are interpreted as right actions on the
coset\footnote{%
For some earlier work in this direction, see for example \cite%
{McArthur:1999dy}}.

The general variation of the LI MC one-forms can be described only in terms
of the structure constants $\left( {f^{A}}_{BC}\right) $, the LI MC forms
and the variation of the Goldstone fields themselves\footnote{%
In the case of kappa transformations of superbranes, see for example \cite%
{Bergshoeff:1987qx} \cite{Bergshoeff:1996tu}}
\begin{equation}
\delta L^{A}=d[\delta Z^{A}]+{f^{A}}_{BC}L^{C}\left[ \delta Z^{B}\right] ,
\label{deltaL}
\end{equation}%
where $[\delta Z^{A}]$ is $L^{A}$ in which $dZ^{M}$ is replaced by $\delta
Z^{M}$
\begin{equation}
\lbrack \delta Z^{A}]=\delta Z^{M}{L_{M}}^{A}\qquad \text{{for}}\qquad
L^{A}=dZ^{M}{L_{M}}^{A}  \label{deltabZ}
\end{equation}%
The crucial point is the relation between $[\delta Z^{A}]$ and the right {%
transformation on the group element}
\begin{equation}
g(Z^{M})\rightarrow g(Z^{M})e^{i\epsilon ^{A}G_{A}}=g(Z^{M}+\delta
_{R}Z^{M})\qquad \lbrack \delta _{R}Z^{A}]=\epsilon ^{A},  \label{rt}
\end{equation}%
where $\delta _{R}Z$ now refers to the right {action} of the Goldstone field
$Z$. After the pullback is taken on the LI MC one-forms, the $\epsilon $
parameter can be made local, $\epsilon \rightarrow \epsilon (\tau )$. Using (%
\ref{deltaL}), the LI MC variations can be computed:
\begin{eqnarray}
\delta _{R}L^{H} &=&d[\delta _{R}t]+L^{D}[\delta _{R}t]-L^{H}[\delta _{R}z]+{%
i}L^{Q}[\delta _{R}{\eta ^{\dagger }}]-i[\delta _{R}\eta ]{L^{Q^{\dagger }}}
\\
\delta _{R}L^{K} &=&d[\delta _{R}w]-L^{D}[\delta _{R}w]+L^{K}[\delta _{R}z]+{%
i}L^{S}[\delta _{R}{\ \lambda ^{\dagger }}]-i[\delta _{R}\lambda ]{%
L^{S^{\dagger }}} \\
\delta _{R}L^{D} &=&d[\delta _{R}z]+2L^{K}[\delta _{R}t]-2L^{H}[\delta
_{R}w] +L^{Q}[\delta _{R}{\ \lambda ^{\dagger }}]+[\delta _{R}\lambda ]{%
L^{Q^{\dagger }}}-[\delta _{R}\eta ]{L^{S^{\dagger }}}-L^{S}[\delta _{R}{%
\eta ^{\dagger }}]  \notag \\
&& \\
\delta _{R}L^{a} &=&d[\delta _{R}\theta ^{a}]+\epsilon ^{abc}L^{c}[\delta
_{R}\theta ^{b}]-i\left( L^{Q}\sigma ^{a}[\delta _{R}{\lambda ^{\dagger }}%
]-[\delta _{R}\lambda ]\sigma ^{a}{L^{Q^{\dagger }}}-[\delta _{R}\eta
]\sigma ^{a}{\ L^{S^{\dagger }}}\right)  \notag \\
&&-iL^{S}\sigma ^{a}[\delta _{R}{\eta ^{\dagger }}]
\end{eqnarray}%
The invariance of the action --modulo surface terms-- under the above
variations requires
\begin{eqnarray}
\begin{pmatrix}
b_{H} & 0 & -b_{K} \\
0 & b_{H} & 2b_{D} \\
2b_{D} & b_{K} & 0%
\end{pmatrix}%
\begin{pmatrix}
\left[ \delta _{R}t\right] \\
\left[ \delta _{R}z\right] \\
\left[ \delta _{R}w\right]%
\end{pmatrix}
&=&0  \label{diffcon} \\
&&  \notag \\
b^{a}\epsilon _{abc}[\delta _{R}\theta ^{b}] &=&0  \label{u1con} \\
&&  \notag \\
\begin{pmatrix}
\left[ \delta _{R}\eta \right] ,\left[ \delta _{R}\lambda \right]%
\end{pmatrix}%
\begin{pmatrix}
b_{H} & -ib_{D}-b^{a}\sigma ^{a}, \\
ib_{D}-b^{a}\sigma ^{a} & b_{K}%
\end{pmatrix}
&=&0.  \label{Kmat}
\end{eqnarray}%
Provided the determinant of the system vanishes, this homogeneous equations
have non-trivial solutions for $[\delta _{R}Z^{M}]$ corresponding to the
different local invariances.

Since the determinant appearing in eq (\ref{diffcon}) vanishes, there is a
non-trivial solution
\begin{equation}
\left[ \delta _{R}t\right] =\epsilon (\tau ),\quad \left[ \delta _{R}z\right]
=-2\frac{b_{D}}{b_{K}}\epsilon (\tau ),\quad \left[ \delta _{R}w\right] =%
\frac{b_{H}}{b_{K}}\epsilon (\tau ),\quad {\text{\textrm{and others}}=0},
\label{Htrans}
\end{equation}%
where $\epsilon (\tau )$ is an arbitrary function. We will refer to this
transformation as $T$ symmetry.

The (\ref{u1con}) is the local $U(1)$ transformation
\begin{equation}
\left[ \delta_R\theta ^{a}\right] =b^{a}\alpha (\tau ),\qquad {\text{\textrm{%
\ and others}}=0},  \label{U1trans}
\end{equation}%
where $\alpha (\tau )$ is an arbitrary function.

The $T$ and $U(1)$ symmetries are present for any non-vanishing value of the
coupling constants. This implies that the number of physical bosonic degrees
of freedom described by the action (\ref{SUlagr}) is two, therefore it {is
not describing} the most general motion of the test particle in the near
horizon of geometry of a $\mathcal{N}=2$ charged four-dimensional extremal
black hole, that has three bosonic degrees of freedom.

The number of linearly realized worldline supersymmetries of the lagrangian
is related to the rank of the matrix in (\ref{Kmat}). When $b_Hb_K\neq
b_D^2+b^{a}b^a$ the $4\times 4 $ matrix in (\ref{Kmat}) has the maximal rank
and (\ref{Kmat}) only has trivial solution $[\delta_R\eta
]=[\delta_R\lambda]=0 $. In this case the system has no local fermionic
symmetry and all supersymmetries are broken (non-BPS particle).

If
\begin{equation}
b_{H}b_{K}-b_{D}^{2}=b^{a}b^{a},  \label{kappacond}
\end{equation}%
the rank of the matrix (\ref{Kmat}) is $2$ and the number of linearly
realized supersymmetries of the worldline is 4 (BPS particle). This relation
implies the equality between the Casimir invariants of the SU(2) and SO(1,2)
sectors.

The action acquires a new local symmetry, the so-called $\kappa $ symmetry.
The corresponding non-trivial solution is
\begin{equation}
\lbrack \delta _{R}\eta ]=\kappa _{\eta }(\tau ),\qquad \lbrack \delta
_{R}\lambda ]=\kappa _{\eta }(\tau )(\frac{ib_{D}}{b_{K}}+\frac{b^{a}\sigma
_{a}}{b_{K}}),\qquad  \label{kaptr1}
\end{equation}%
and other bosonic $[\delta _{R}Z^{A}]$ are zero,
\begin{equation}
\left[ \delta _{R}t\right] =\left[ \delta _{R}z\right] =\left[ \delta
_{R}\omega \right] =\left[ \delta _{R}\theta ^{a}\right] =0,
\label{bosonkap}
\end{equation}%
where $\kappa _{\eta }^{i}(\tau )$ is a $SU(2)$ doublet arbitrary
Grassman-valued function of $\tau $.

Following \cite{Gomis:2006wu} we can construct the generators of the local
algebra. In our context the local symmetries $T$, $U(1)$ and $\kappa $, are
\begin{eqnarray}
T &=&H-2\frac{b_{D}}{b_{K}}D+\frac{b_{H}}{b_{K}}K-2\frac{b^{a}}{b_{K}}J_{a},
\\
B &=&b^{a}J_{a}, \\
\tilde{Q}^{i} &=&Q^{i}+S^{j}\left( -\frac{ib_{D}}{b_{K}}{\delta _{j}}^{i}+%
\frac{b^{a}}{b_{K}}{(\sigma _{a})_{j}}^{i}\right) \\
\tilde{Q}_{i}^{\dagger } &=&Q_{i}^{\dagger }+\left( \frac{ib_{D}}{b_{K}}{%
\delta _{i}}^{j}+\frac{b^{a}}{b_{K}}{(\sigma _{a})_{i}}^{j}\right)
S_{j}^{\dagger }.
\end{eqnarray}

In the case of (\ref{kappacond}) they generate unbroken symmetry of the
Lagrangian (\ref{SUlagr}) and form a subalgebra of the $psu(1,1|2)$,
\begin{eqnarray}
\left[ \tilde{Q}^{i},\tilde{Q}_{j}^{\dagger }\right] _{+} &=&{\delta ^{i}}%
_{j}T,\qquad \left[ \tilde{Q}^{i},\tilde{Q}^{j}\right] _{+}=\left[ \tilde{Q}%
^{i},T\right] =0, \\
\left[ B,\tilde{Q}^{i}\right] &=&\frac{1}{2}\tilde{Q}^{j}{(b^{a}\sigma
_{a})_{j}}^{i},\qquad \left[ B,T\right] =0.
\end{eqnarray}

The diffeomorphism invariance, $\tau \rightarrow \tau ^{\prime }(\tau )$ is
not independent of the local symmetries previously discussed. Moreover, when
the condition (\ref{kappacond}) for $\kappa $ symmetry is satisfied,
diffeomorphims are equivalent to linear combinations of the local symmetries
obtained from the right translations, with parameters chosen in terms of $%
\delta\tau=\varepsilon (\tau )$ as
\begin{equation}
\epsilon (\tau )=(L^{H})^{\ast }\varepsilon (\tau ),\qquad \alpha (\tau )=%
\frac{(b^{b}L^{b})^{\ast }}{b^{c}b^{c}}\varepsilon (\tau ),\qquad \kappa
_{\eta }(\tau )s(\theta )=(L^{Q})^{\ast }\varepsilon (\tau ).
\end{equation}%
In the non-BPS case there is no kappa transformation.

In the Appendix C it is shown that these combinations of the local
transformations and diffeomorphisms differ by trivial variations, \textit{%
i.e.} (graded)anti-symmetric combinations of the equations of motion.

\subsection{Kappa symmetry}

It has been shown that if the constants of the Lagrangian satisfy (\ref%
{kappacond}), the action is invariant under the kappa transformations. The
transformation of the fields around the configuration $\eta =\eta ^{\dagger
}=0,$\footnote{%
The transformation for a general configuration is rather complicated. We
give it in the $OSP(2\left\vert 2\right. )$ case in the appendix D.} is
obtained from (\ref{kaptr1}) and (\ref{bosonkap}) as
\begin{equation}
\delta _{\kappa }\eta |_{\eta =\eta ^{\dagger }=0}={\kappa }_{\eta }{%
s(\theta )^{-1}},\qquad \delta _{\kappa }\eta ^{\dagger }|_{\eta =\eta
^{\dagger }=0}=\kappa _{\eta ^{\dagger }}{s(\theta )^{-1}}.
\label{kappatrans3}
\end{equation}%
Where $s(\theta )$ is the spin one half representation of the $SU(2)$ group,
by redefinition of the parameter $\kappa $, $s(\theta )$ can be reabsorbed.
Then it follows that in any neighborhood of ${\eta =\eta ^{\dagger }=0}$ the
gauge slice
\begin{equation}
{\eta =\eta ^{\dagger }=0}
\end{equation}%
is accessible. In this gauge the remaining coordinates transform as
\begin{eqnarray}
\delta _{\kappa }\lambda |_{\eta =\eta ^{\dagger }=0} &=&\kappa _{\eta }-%
\frac{1}{2}{\kappa }_{\eta }(\lambda {\lambda ^{\dagger }})-\frac{1}{2}%
\lambda ({\kappa }_{\eta }{\lambda ^{\dagger }}-\lambda \kappa _{\eta
^{\dagger }}) \\
\delta _{\kappa }t|_{\eta =\eta ^{\dagger }=0} &=&0 \\
\delta _{\kappa }z|_{\eta =\eta ^{\dagger }=0} &=&-(\lambda \delta _{\kappa
}\eta ^{\dagger }+\delta _{\kappa }\eta {\lambda ^{\dagger }}) \\
\delta _{\kappa }\omega |_{\eta =\eta ^{\dagger }=0} &=&-\omega (\lambda
\delta _{\kappa }\eta ^{\dagger }+\delta _{\kappa }\eta {\lambda ^{\dagger }}%
)+\frac{i}{2}(\lambda \delta _{\kappa }{\lambda ^{\dagger }}-\delta _{\kappa
}\lambda {\lambda ^{\dagger }})+\frac{i}{2}(\lambda \delta _{\kappa }{\eta
^{\dagger }}-\delta _{\kappa }\eta {\lambda ^{\dagger }})(\lambda {\lambda
^{\dagger }})  \notag \\
&& \\
\delta _{\kappa }\theta ^{a}|_{\eta =\eta ^{\dagger }=0} &=&-i(\lambda
\sigma _{b}\delta _{\kappa }{\eta ^{\dagger }}-\delta _{\kappa }{\eta }%
\sigma _{b}{\lambda ^{\dagger }})\left( \mathbf{R}_{b}^{a}\right) ^{-1}
\end{eqnarray}

As can be seen from the previous results, when the kappa condition (\ref%
{kappacond}) is satisfied, it is possible to gauge away half of the
fermions. In the next section diffeomorphism and kappa symmetry are further
fixed, residual transformations found and BPS states obtained.

\subsection{Gauge fixed lagrangian and residual global transformations}

The kappa symmetry can be used to further simplify the form of the
lagrangian. In fact, imposing (\ref{kappacond}), and setting $\eta =\eta
^{\dagger }=0$ and {the static gauge } $t=\tau $, the action (\ref{suinv})
becomes
\begin{equation}
\mathcal{S}=\int dt\;b_{K}\left[ \frac{\dot{q}^{2}}{2}-\frac{2}{q^{2}}\left(
\frac{1}{4}(\lambda \lambda ^{\dagger })^{2}-(\lambda \sigma _{a}\lambda
^{\dagger })\mathcal{S}_{ab}\frac{b_{b}}{b_{K}}+\frac{b_{a}b_{a}}{b_{K}^{2}}%
\right) -\frac{i}{2}(\lambda \dot{\lambda}^{\dagger }-\dot{\lambda}\lambda
^{\dagger })\right] +\left( L_{a}^{0}\right) ^{\ast }b_{a},  \label{fixed}
\end{equation}%
where, in this gauge,
\begin{equation}
q=\sqrt{2}e^{\frac{z}{2}}.
\end{equation}

Moreover the coupling constant of $q^{-2}$ computed in \cite{Claus:1998ts}
for the kappa-symmetric particle $(e=m)$ is exactly reproduced.\newline

\begin{equation}
\frac{g}{2}=4\frac{b_{a}b_{a}}{b_{K}}=4\frac{J^{2}}{m}  \label{g}
\end{equation}

As it was previously pointed out, $\theta ^{3}$ is non dynamical, its
elimination reduces (\ref{fixed}) to%
\begin{eqnarray}
\mathcal{L} &=&b_{K}\frac{\dot{q}^{2}}{2}-\frac{2}{{q}^{2}b_{K}}\left(
b_{K}^{2}\frac{1}{4}(\lambda \lambda ^{\dagger })^{2}+\frac{b_{a}b_{a}}{%
b_{K}^{2}}\right) -\frac{i}{2}b_{K}(\lambda \dot{\lambda}^{\dagger }-\dot{%
\lambda}\lambda ^{\dagger })  \notag \\
&&+\sqrt{b_{1}^{2}+b_{2}^{2}}\sqrt{\left( \dot{\theta}_{1}\cos \theta
_{2}+j\cdot e_{1}\right) ^{2}+\left( \dot{\theta}_{2}+j\cdot e_{2}\right)
^{2}}  \notag \\
&&+b_{3}\left( -\dot{\theta}_{1}\sin \theta _{2}+j\cdot e_{3}\right) ,
\label{last}
\end{eqnarray}%
where $j_{a}=2\frac{(\lambda \sigma _{a}\lambda ^{\dagger })}{q^{2}}$ and
the orthonormal basis $e_{a}$ is given by

\begin{equation}
e_{1}=\left(
\begin{array}{c}
\cos \theta _{2} \\
\sin \theta _{1}\sin \theta _{2} \\
\cos \theta _{1}\sin \theta _{2}%
\end{array}%
\right) \qquad e_{2}=\left(
\begin{array}{c}
0 \\
\cos \theta _{1} \\
-\sin \theta _{1}%
\end{array}%
\right) \qquad e_{3}=\left(
\begin{array}{c}
-\sin \theta _{2} \\
\sin \theta _{1}\cos \theta _{2} \\
\cos \theta _{1}\cos \theta _{2}%
\end{array}%
\right) .
\end{equation}

It must be noted that action (\ref{last}) still has the U(1) gauge
invariance (\ref{U1trans}), $\delta \theta ^{b}=\alpha (\tau )b_{a} \left(
\mathbf{L}_{a}^{b}\right) ^{-1},$ which after $\theta_3$ is eliminated
becomes
\begin{equation}
\delta \theta_{1}=\tilde\alpha (\tau ) (\dot\theta_1+j\cdot
e_1/\cos\theta_2),\qquad \delta \theta_{2}=\tilde\alpha (\tau )
(\dot\theta_2+j\cdot e_2),  \label{U1trans2}
\end{equation}%
where
\begin{equation}
\tilde\alpha(\tau )=\frac{\sqrt{b_{1}^2+b_2^2} }{ \sqrt{(\dot\theta_1+j\cdot
e_2/\cos\theta_2)^2+ (\dot\theta_2+j\cdot e_2)^2}}\alpha (\tau ),
\end{equation}%
and $\delta Z^{M}=0$ for other fields.

The gauge fixing has changed the form of the global transformations. This is
because local transformations must be used in order to respect the gauge
slice previously chosen. This means that local compensating transformations
must be introduced. It is straightforward to show that they are given by
\begin{eqnarray}
\delta ^{\ast }\left( t-\tau \right) |_{{\eta =\eta ^{\dagger }=0};t=\tau
}=\left( \varepsilon \dot{t}+\delta _{\kappa }t+\delta _{G}t\right) |_{{\eta
=\eta ^{\dagger }=0};t=\tau }=0 &\Longrightarrow &\varepsilon =-\delta _{G}t,
\label{llave2} \\
\delta ^{\ast }\eta |_{{\eta =\eta ^{\dagger }=0};t=\tau }=\left(
\varepsilon \dot{\eta}+\delta _{\kappa }\eta +\delta _{G}\eta \right) |_{{%
\eta =\eta ^{\dagger }=0};t=\tau }=0 &\Longrightarrow &\kappa _{\eta
}=-\delta _{G}\eta \\
\delta ^{\ast }\eta ^{\dagger }|_{{\eta =\eta ^{\dagger }=0};t=\tau }=\left(
\varepsilon \dot{\eta}^{\dagger }+\delta _{\kappa }\eta ^{\dagger }+\delta
_{G}\eta ^{\dagger }\right) |_{{\eta =\eta ^{\dagger }=0};t=\tau }=0
&\Longrightarrow &\kappa _{\eta ^{\dagger }}=-\delta _{G}\eta ^{\dagger },
\label{llave3}
\end{eqnarray}%
where $\delta _{G}$ stands for any global $SU(\left. 1,1\right\vert 2)$
transformation. The residual transformations for the remaining coordinates $%
\delta ^{\ast }$ are defined in the same way {\ as} the former variations,
but with the local parameters given by (\ref{llave2}-\ref{llave3}),
\begin{eqnarray}
\delta _{H}^{\ast }q &=&\epsilon _{H}\dot{q}\qquad \delta _{H}^{\ast
}\lambda =\epsilon _{H}\dot{\lambda}\qquad \delta _{H}^{\ast }\theta
^{a}=\epsilon _{H}\dot{\theta}^{a} \\
\delta _{D}^{\ast }q &=&\epsilon _{D}\left( \frac{q}{2}-t\dot{q}\right)
\qquad \delta _{D}^{\ast }\lambda =-t\epsilon _{D}\dot{\lambda}\qquad \delta
_{D}^{\ast }\theta ^{a}=-t\epsilon _{D}\dot{\theta}^{a} \\
\delta _{K}^{\ast }q &=&-t\epsilon _{K}q+t^{2}\epsilon _{K}\dot{q}\qquad
\delta _{K}^{\ast }\lambda =t^{2}\epsilon _{K}\dot{\lambda}\qquad \delta
_{K}^{\ast }\theta ^{a}=t^{2}\epsilon _{K}\dot{\theta}^{a} \\
\delta _{a}^{\ast }q &=&0\qquad \delta _{a}^{\ast }\lambda =-\frac{i}{2}%
\lambda \sigma _{a}\epsilon \qquad \delta _{a}\theta ^{b}=\mathbf{R}%
_{ab}^{-1}\epsilon \\
\delta _{Q_{i}}^{\ast }q &=&\frac{1}{\sqrt{2}}\epsilon _{Q}{\lambda
^{\dagger }}\qquad \delta _{Q_{i}}^{\ast }\lambda _{k}^{\dagger }=\frac{1}{%
\sqrt{2}q}\lambda _{k}^{\dagger }\lambda _{i}^{\dagger }\epsilon _{Q}\qquad
\delta _{Q_{i}}^{\ast }\theta ^{a}=-i\epsilon _{Q}\left( \sigma _{b}{\lambda
^{\dagger }}\right) _{i}\mathbf{R}_{ba}^{-1}\frac{\sqrt{2}}{q}  \notag \\
\delta _{Q_{i}}^{\ast }\lambda _{k} &=&\left( i\left( \frac{-\dot{\widetilde{%
q}}q}{2}\right) \delta _{ik}-\frac{b_{a}}{b_{K}}\left( {\sigma }_{b}\right)
_{i}^{k}\mathcal{S}_{ba}+\frac{1}{2}{\delta }_{ik}\lambda {\lambda ^{\dagger
}}-\frac{1}{2}\lambda _{k}{\lambda _{i}^{\dagger }}\right) \epsilon _{Q}%
\frac{\sqrt{2}}{q} \\
\delta _{S_{i}}^{\ast }q &=&it\frac{1}{\sqrt{2}}{\lambda _{i}^{\dagger }}%
\epsilon _{S}\qquad \delta _{S_{i}}^{\ast }\lambda _{k}^{\dagger }=-\frac{it%
}{\sqrt{2}q}{\lambda _{k}^{\dagger }\lambda _{i}^{\dagger }}\epsilon
_{S}\qquad \delta _{S_{i}}^{\ast }\theta ^{a}=-\frac{t\sqrt{2}}{q}\epsilon
_{S}\left( \sigma _{b}{\lambda ^{\dagger }}\right) ^{i}\mathbf{R}_{ba}^{-1}
\notag \\
\delta _{S_{i}}^{\ast }\lambda _{k} &=&\bigskip \left( \left( \frac{q^{2}}{2}%
-\frac{\dot{qq}}{2}t\right) \delta _{ik}+it\frac{b_{a}}{b_{K}}\left( {\sigma
}_{b}\right) _{i}^{k}\mathcal{S}_{ba}-it\frac{1}{2}\delta _{ik}(\lambda {%
\lambda ^{\dagger }})+\frac{1}{2}it\lambda _{k}{\lambda _{i}^{\dagger }}%
)\right) \frac{\sqrt{2}}{q}\epsilon _{S}
\end{eqnarray}%
\medskip

To study the existence of BPS states, the residual transformation of the
fermions under $Q$'s and $S$'s are considered. In this way, two BPS
equations arise:
\begin{eqnarray}
\delta _{Q_{i}}^{\ast }\lambda _{k} &=&\delta _{Q_{i}^{\dagger }}^{\ast
}\lambda _{k}^{\dagger }=0\Longrightarrow \left( \frac{\dot{q}q}{2}\right)
^{2}+\frac{b_{a}b_{a}}{b_{K}^{2}}=0,  \label{BPS1} \\
\delta _{S_{i}}^{\ast }\lambda _{k} &=&\delta _{S_{i}^{\dagger }}^{\ast
}\lambda _{k}^{\dagger }=0\Longrightarrow \left( \frac{q^{2}}{2}-t\frac{\dot{%
q}q}{2}\right) ^{2}+t^{2}\frac{b_{a}b_{a}}{b_{K}^{2}}=0,  \label{BPS2}
\end{eqnarray}%
As both of them are the sum of two positive terms, a necessary condition for
the existence of BPS states is $b_{a}=0$. Then, by Eq. (\ref{g}), the
coupling constant vanishes and the system is just the free particle. Then,
the $\dot{q}=0$ configuration, (\ref{BPS1}), saturates the bound of the free
particle Hamiltonian $H$, meanwhile the $\dot{q}=\frac{q}{t}$ configuration,
(\ref{BPS2}),\ saturates the bound of the free particle special conformal
transformation generator $K$. So the existence of non trivial 1/2 BPS states
is ruled out.

\subsection{Bosonic Motions}

Let us now study the bosonic trajectories of our model. If we fix the
diffeomorphism by taking the gauge $\dot{t}=1$ the Lagrangian (\ref{blag1})
becomes
\begin{eqnarray}
\mathcal{S} &=&\int d\tau \left[ m\;\frac{\dot{q}^{2}}{2}-\frac{2}{mq^{2}}%
\left( {b_{H}b_{K}-b_{D}^{2}}\right) +\sqrt{{b_{1}^{2}+b_{2}^{2}}}\sqrt{\dot{%
\phi}^{2}\sin ^{2}\theta +\dot{\theta}^{2}}+{eg}\;\left( -\dot{\phi}\cos
\theta \right) \right] ,  \notag  \label{blag3} \\
&&
\end{eqnarray}%
where $\;\phi =\theta _{1},\;\theta =\frac{\pi }{2}-\theta _{2}.$ The first
class constraint associated to the $U(1)$ gauge invariance of the Lagrangian
is,
\begin{equation}  \label{gk}
\Psi =\frac{1}{2}\left[ p_{\theta}^{2}+ \left(\frac{p_{\phi }+eg\cos \theta }{\sin
\theta }\right)^{2}-(b_{1}^{2}+b_{2}^{2})\right] \sim 0,
\end{equation}%
where $p_{q},p_{\theta},p_{\phi }$ are the canonical momenta associated to the
coordinates $q,\theta ,\phi $. The Dirac Hamiltonian is
\begin{equation}
H=\frac{p_{q}^{2}}{2m}+\frac{2}{q^{2}}\left( \frac{b_{H}b_{K}-b_{D}^{2}}{%
b_{K}}\right) +\Lambda \Psi ,
\end{equation}%
where $\Lambda $ is an arbitrary function of $t$. In the presence of a
monopole background the conserved angular momentum is
\begin{equation}
\mathbf{J}=p_{\theta}\mathbf{e}_{\phi }-\left(\frac{p_{\phi }+eg\cos \theta }{\sin
\theta }\right)\mathbf{e}_{\theta}-(eg)\mathbf{e}_{r},
\end{equation}%
where $\mathbf{e}_{r},\mathbf{e}_{\theta},\mathbf{e}_{\phi }$ are the othonormal
unit vectors in the polar coordinates. The constraint (\ref{gk}) means that the
value of $\mathbf{J}^{2}$ is fixed by parameters of the lagrangian as
\begin{equation}
\mathbf{J}^{2}=p_{\theta}^{2}+\left(\frac{p_{\phi }+eg\cos \theta }{\sin
\theta}\right)^{2}+(eg)^{2}\sim b_{a}^{2}.
\label{U1con3}
\end{equation}

We have considered the kappa invariant case when (\ref{kappacond}) is
satisfied. Furthermore, in a gauge where the arbitrary function $\Lambda $
appears as
%ing in the Dirac hamiltonian \bref{hamil} is fixed when a gauge fixing
%condition $\chi=0$ is imposed. We do not specify its form but we assume
%the gauge condition is such that it determines the $\Lam$ as
\begin{equation}
\Lambda =\frac{4}{mq^{2}}
\end{equation}%
the Hamiltonian becomes
\begin{equation}
H^{\ast }=\frac{p_{q}^{2}}{2m}+\frac{2}{mq^{2}}\;\mathbf{J}^{2}=\frac{%
p_{q}^{2}}{2m}+\frac{2}{mq^{2}}\left( p_{\theta}^{2}+(\frac{p_{\phi }+eg\cos
\theta }{\sin \theta })^{2}+(eg)^{2}\right) .
\end{equation}%
The corresponding lagrangian is
\begin{equation}
L^{\ast }=\frac{m}{2}\dot{q}^{2}+\frac{mq^{2}}{8}\left( \dot{\theta}^{2}+%
\dot{\phi}^{2}\sin ^{2}\theta \right) -2\frac{(eg)^{2}}{mq^{2}}-(eg)\dot{\phi%
}\cos \theta .
\label{Leff}
\end{equation}%
The Lagrangian (\ref{Leff}) agrees with the bosonic part of the $D(2,1,\alpha
=-1)$ superconformal lagrangian considered in \cite{Ivanov:2002tb}. It
coincides also with that of the D0 particle on a black hole attractor
\cite{Gaiotto:2004ij} up to second order expansion in derivatives.

Although the form of the lagrangians coincides, the physical content of these
models is different. The lagrangian (\ref{Leff}) has associated the constraint
(\ref{gk});
\begin{equation}
\Psi ^{\ast }=\frac{1}{2}\left[ (\frac{mq^{2}}{4})^{2}(\dot{\theta}^{2}+\dot{%
\phi}^{2}\sin ^{2}\theta )-(b_{1}^{2}+b_{2}^{2})\right] \sim 0.
\label{U1con4}
\end{equation}%
In both cases the trajectories of the particle are on a two dimensional cone.
However, in our case the total angular momentum squared is constrained by
(\ref{U1con4}), it follows that in terms of the parameters of our lagrangian
the opening angle of the cone is fixed as $\tan
^{-1}(\frac{\sqrt{b_{1}^{2}+b_{2}^{2}}}{b_{3}})$.

\section{Covariant AdS parametrization}

At the quantum level, the conformal mechanics has no ground state associated
to the hamiltonian $H$. The wave function spreads out to spatial infinity.
The authors of \cite{deAlfaro:1976je} suggest that one should consider the
eigenstates of the compact operator $P_{0}=\frac{1}{2}(H+K)$ which has a
discrete spectrum of normalizable eigenstates. From the perspective of the
particle motion near the black hole it corresponds to a different choice of
time \cite{Claus:1998ts}. In fact the conjugate variable to $P_{0}$ is the
global time of $AdS_{2}$ and can describe the motion of the particle
entering through the horizon, instead the time conjugate to $H$ only
describe the motion of the particle outside of the horizon. Therefore it is
also natural to study the dynamics of the superconformal particle using the
new basis, that we call \emph{AdS basis}. In our approach this implies a new
parametrization of the coset, that we take
\begin{equation}
g\;=\;g_{0}^{AdS_{2}}\;g_{0}^{S^{2}}\;{e^{i(Q\bar{\eta}+\eta \bar{Q})}}\;{%
e^{i(S\bar{\lambda}+\lambda \bar{S})}}\;e^{iM_{01}y}\;e^{iJ_{3}{y^{\prime }}%
},  \label{cosetAdS}
\end{equation}%
where
\begin{equation}
g_{0}^{AdS_{2}}=e^{iP_{0}x^{0}}e^{iP_{1}x^{1}},\qquad g_{0}^{S^{2}}=e^{iJ_{1}%
{\theta }^{1}}e^{iJ_{2}{\theta }^{2}}
\end{equation}%
and the $AdS_{2}$ generators $P_{0},P_{1},M_{01}$ are related to the
conformal ones by
\begin{equation}
P_{0}=\frac{H+K}{2},\qquad P_{1}=D,\qquad M_{01}=\frac{H-K}{2}.
\end{equation}%
The MC one form is
\begin{equation}
\Omega =L^{P_{\mu }}P_{\mu
}+L^{M_{01}}M_{01}+L^{J_{a}}J_{a}+L^{J_{3}}J_{3}+QL^{\dagger
Q}+L^{Q}Q^{\dagger }+SL^{\dagger S}+L^{S}S^{\dagger }.  \label{mcleftAdS}
\end{equation}%
where $\mu =0,1$ and $a=1,2$. The invariant particle Lagrangian is a sum of
bosonic forms
\begin{equation}
\mathcal{L}=L^{P_{\mu }}b_{P_{\mu
}}+L^{M_{01}}b_{M_{01}}+L^{J_{a}}b_{a}+L^{J_{3}}b_{3}.  \label{LagAdS}
\end{equation}%
In (\ref{cosetAdS}) we have put $e^{iM_{01}{y}},\;e^{iJ_{01}y^{\prime }}$ at
the right so that $dy$ and $dy^{\prime }$ terms appear in the lagrangian in
total derivative forms and can be omitted. The Lagrangian is written as
\begin{equation}
\mathcal{L}=(A\;\sinh y+B\;\cosh y+C)+(A^{\prime }\;\sin y^{\prime
}+B^{\prime }\;\cos y^{\prime }+C^{\prime }).
\end{equation}%
where
\begin{eqnarray}
A &=&b_{P_{1}}L_{1}^{P_{0}}+b_{P_{0}}L_{1}^{P_{1}},\qquad
B=b_{P_{0}}L_{1}^{P_{0}}+b_{P_{1}}L_{1}^{P_{1}},\qquad
C=b_{M_{01}}L_{1}^{M_{01}}  \notag \\
A^{\prime } &=&b_{2}L_{1}^{J_{1}}-b_{1}L_{1}^{J_{2}},\qquad B^{\prime
}=b_{1}L_{1}^{J_{1}}+b_{2}L_{1}^{J_{2}},\qquad C^{\prime }=b_{3}L_{1}^{J_{3}}
\label{ABC}
\end{eqnarray}%
where the explicit forms of $L_{1}^{A}$'s are given in the Appendix A. The
Goldstone fields $y$ and $y^{\prime }$ are non-dynamical variables and can
be eliminated using their equations of motion and
\begin{eqnarray}
\mathcal{L} &=&-\sqrt{b_{P_{0}}^{2}-b_{P_{1}}^{2}}\;\sqrt{%
(L_{1}^{P_{0}})^{2}-{(L_{1}^{P_{1}})}^{2}}+b_{M_{01}}\;L_{1}^{M_{01}}  \notag
\\
&&-\sqrt{b_{1}^{2}+b_{2}^{2}}\;\sqrt{(L_{1}^{J_{1}})^{2}+{(L_{1}^{J_{2}})}%
^{2}}+b_{3}\;L_{1}^{J_{3}}.  \label{LagAdS2}
\end{eqnarray}%
As the previous discussions in section 4 the action from (\ref{LagAdS2}) is
invariant under two bosonic local symmetries, diffeomorphism and U(1). It is
also invariant under the kappa symmetry if the coefficients of the
Lagrangian are verifying
\begin{equation}
b_{P_{0}}^{2}-b_{P_{1}}^{2}-b_{M_{01}}^{2}=b_{a}b_{a}.
\end{equation}%
which is corresponding to (\ref{kappacond}), $%
b_{H}b_{K}-b_{D}^{2}=b_{a}b_{a} $.

As in the \emph{conformal }basis this relation implies and equality between
the Casimir invariants of $SU(2)$ and $SU(1,1)$. The two WZ terms represents
the coupling to the electromganetic field.

The lagrangian (\ref{LagAdS2}), where the fermions have been set to zero,
\begin{eqnarray}
\mathcal{L} &=&-\sqrt{b_{P_{0}}^{2}-b_{P_{1}}^{2}}\; \sqrt{(dx^0\cosh\frac{%
x^1}{R})^{2}-{(dx^1)}^{2}} +b_{M_{01}}\;{\frac{dx^0}{R}\sinh\frac{x^1}{R}}
\notag \\
&&-\sqrt{b_{1}^{2}+b_{2}^{2}}\;\sqrt{(d\theta^1\cos\theta^2)^{2}+ {%
(d\theta^2)}^{2}}-b_{3}\;d\theta^1\sin\theta^2,  \label{LagAdS2b}
\end{eqnarray}%
does not reproduce the motion of a relativistic particle in $AdS_{2}\times
S_{2}$, because the lagrangian here has two square roots which is not
equivalent to the lagrangian studied in references \cite{Billo:1999ip}, \cite%
{Zhou:1999sm}, \cite{Gaiotto:2004ij},\cite{Hatsuda:2003er}. The two systems
have different numbers of degrees of freedom since they possess different
bosonic gauge symmetries. A similar effect occurs in the conformal basis due
to the appearance of two gauge symmetries, diffeomorphisms and $U(1)$
transformations. In the $D0$ brane lagrangian, instead, there are only
diffeomorphisms. Since we have interpreted the gauge transformations as
induced by the right action on the coset by unbroken translation \cite%
{Gomis:2006wu}, it means that there are two unbroken translations given by $%
P_{0}$ and the $b^{a}J_{a}$ in the present case.

\section{Discussion and Outlook}

The BPS and non BPS dynamics of a superconformal particle has been
constructed, using \textit{only} the method of non-linear realization
without resorting to superfields or requiring further constraints \cite%
{Gomis:2006xw}. The coset $PSU(1,1|2)$\ had been considered, as in \cite%
{Ivanov:1988it} \cite{deAzcarraga:1998ni}. The particle action contains six
couplings constants and is invariant under two set of bosonic gauge
symmetries, diffeomorphisms and $U(1)$ gauge transformations. When the
condition on the coupling constants (\ref{kappacond}) is verified, the
action becomes also kappa symmetric. This relation implies the equality
between the Casimir operators of the $SU(2)$ and the $SU(1,1)$ sectors.
Following reference \cite{Gomis:2006wu} these gauge symmetries can be
interpreted as generated by the unbroken ``translations" via the right
action. Furthermore, the algebra verified by the generators of gauge
transformations was found.

The description of the dynamics has been done in two different bases or
parametrizations of the coset: the conformal basis and the AdS basis. In
both cases the kappa-symmetric and non kappa-symmetric models can be viewed
as describing the equatorial motion of a particle near the horizon of a $%
\mathcal{N}=2$ charged four-dimensional extremal black hole. It turns that the
particle has its total angular momentun squared fixed, this value is determined
by the parameters appearing in the lagrangian. They are not describing the
entire three dimensional dynamics of the $D0$ particle.

The analysis of the existence of BPS states shows trough equations (\ref%
{BPS1}) and (\ref{BPS2}) that they only exist in a highly degenerate case of
the conformal mechanics, namely, in the free particle case. A natural
question then arises as to whether it is possible to obtain the lagrangian
of a $D0$ brane from the method of non-linear realization without any extra
geometrical or physical requirements. This point will be addressed in a
future study.

\textbf{Acknowledgements}

We wish to thank to Adolfo de Azc\'arraga, Jaume Gomis, Machiko Hatsuda, Tomas
Ortin, Mikhail Plyushchay, Norisuke Sakai, Paul Townsend, Ricardo Troncoso,
Toine Van Proeyen and Peter West for enlightning discussions, and to Evgeny
Ivanov for calling our attention to some subtleties about the bosonic motion
and for pointing out some pertinent references. A. A. wishes to thanks the warm
hospitality of the Department of Estructura i Constituents de la Mat\`{e}ria of
the Universitat of Barcelona, where part of this project was done; the support
of the project MECESUP USA0108 was essential to achieve this visit. This work
has been supported in part by FONDECYT grants N${{}^{o}}$ 1061291, 1060831,
1040921, from FONDECYT, the European EC-RTN network MRTN-CT-2004-005104, MCYT
FPA 2004-04582-C02-01 and CIRIT GC 2005SGR-00564. Institutional support to the
Centro de Estudios Cient\'{\i}ficos (CECS) from Empresas CMPC is gratefully
acknowledged. CECS is funded in part by grants from the Millennium Science
Initiative, Fundaci\'{o}n Andes, the Tinker Foundation.

\appendix

\section{Appendix: Notation and conventions}

The $psu(1,1|2)$ algebra
\begin{eqnarray}
\left[ H,D\right] &=&iH,\qquad \left[ K,D\right] =-iK,\qquad \left[ H,K%
\right] =2iD, \\
\left[ J_{a},J_{b}\right] &=&i\varepsilon _{abc}J_{c}, \\
\left[ Q^{i},Q_{j}^{\dagger }\right] _{+} &=&{\delta ^{i}}_{j}H,\qquad \left[
S^{i},S{^{\dagger }}_{j}\right] _{+}={\delta ^{i}}_{j}K,\qquad \\
\left[ S{^{\dagger }}_{i},Q^{j}\right] _{+} &=&-{(\sigma _{a})_{i}}%
^{j}J_{a}+i{\delta _{j}}^{i}D,\qquad \left[ Q{^{\dagger }}_{i},S^{j}\right]
_{+}=-{(\sigma _{a})_{i}}^{j}J_{a}-i{\delta _{j}}^{i}D, \\
\left[ D,Q^{i}\right] &=&-\frac{i}{2}Q^{i},\qquad \left[ D,Q{^{\dagger }}_{i}%
\right] =-\frac{i}{2}Q{^{\dagger }}_{i}, \\
\left[ D,S^{i}\right] &=&\frac{i}{2}S^{i},\qquad \left[ D,S{^{\dagger }}_{i}%
\right] =\frac{i}{2}S{^{\dagger }}_{i}, \\
\left[ K,Q^{i}\right] &=&S^{i},\qquad \left[ K,Q{^{\dagger }}_{i}\right] =-S{%
^{\dagger }}_{i}, \\
\left[ H,S^{i}\right] &=&Q^{i},\qquad \left[ H,S{^{\dagger }}_{i}\right] =-Q{%
^{\dagger }}_{i} \\
\left[ J_{a},Q^{i}\right] &=&\frac{1}{2}Q^{j}(\sigma _{a})_{j}^{i},\qquad %
\left[ J_{a},Q{^{\dagger }}_{i}\right] =-\frac{1}{2}(\sigma _{a})_{i}^{j}Q{%
^{\dagger }}_{j},  \label{algebraJQ} \\
\left[ J_{a},S^{i}\right] &=&\frac{1}{2}S^{j}(\sigma _{a})_{j}^{i},\qquad %
\left[ J_{a},S{^{\dagger }}_{i}\right] =-\frac{1}{2}(\sigma _{a})_{i}^{j}S{%
^{\dagger }}_{j}.  \label{algebraJS}
\end{eqnarray}

\newpage
\textbf{Maurer-Cartan forms}

The Maurer-Cartan one-forms are explicitly given by
\begin{eqnarray}
L^{H} &=&L_{H}^{0}+\frac{1}{4}L_{K}^{0}(\eta {\eta ^{\dagger }})^{2}-\frac{i%
}{2}(\eta d{\eta ^{\dagger }}-d\eta {\eta ^{\dagger }}),  \label{1} \\
L^{D} &=&L_{D}^{0}\{1+\frac{1}{2}(\lambda {\eta ^{\dagger }}+\eta {\lambda
^{\dagger }})\}+\frac{i}{2}L_{K}^{0}(\eta {\eta ^{\dagger }})(\lambda {\eta
^{\dagger }}-\eta {\lambda ^{\dagger }})+(\lambda d{\eta ^{\dagger }}+d\eta {%
\lambda ^{\dagger }}),  \label{4} \\
L^{K} &=&L_{K}^{0}\{1+(\lambda {\eta ^{\dagger }}+\eta {\lambda ^{\dagger }}%
)-\frac{1}{2}(\eta \sigma _{a}{\eta ^{\dagger }})(\lambda \sigma _{a}{%
\lambda ^{\dagger }})+\frac{1}{4}(\eta {\eta ^{\dagger }})(\lambda {\lambda
^{\dagger }})(\lambda {\eta ^{\dagger }}+\eta {\lambda ^{\dagger }})\}+\frac{%
1}{4}L_{H}(\lambda {\lambda ^{\dagger }})^{2}  \notag \\
&&-\frac{i}{4}L_{D}^{0}(\lambda {\eta ^{\dagger }}-\eta {\lambda ^{\dagger }}%
)(\lambda \lambda {^{\dagger }})-\frac{i}{2}(\lambda d{\lambda ^{\dagger }}%
-d\lambda {\lambda ^{\dagger }})-\frac{i}{2}(\lambda d{\eta ^{\dagger }}%
-d\eta {\lambda ^{\dagger }})(\lambda {\lambda ^{\dagger }}),  \label{6} \\
L^{J_{b}} &=&L_{J_{b}}^{0}+\left[ i(\lambda \sigma _{a}d{\eta ^{\dagger }}%
-d\eta \sigma _{a}{\lambda ^{\dagger }})+\frac{i}{2}L_{D}^{0}\left( \lambda
\sigma _{a}{\eta ^{\dagger }}-\eta \sigma _{a}{\lambda ^{\dagger }}\right)
\right.  \notag \\
&&\left. +L_{K}^{0}\{-(\eta \sigma _{a}{\eta ^{\dagger }})-\frac{1}{2}(\eta {%
\eta ^{\dagger }})\left( \lambda \sigma _{a}{\eta ^{\dagger }}+\eta \sigma
_{a}{\lambda ^{\dagger }}\right) \}-L_{H}(\lambda \sigma _{a}{\lambda
^{\dagger }})\right] \mathcal{S}_{ab}(\theta ),  \label{3} \\
L^{Q} &=&\left[ d\eta +\frac{1}{2}\;L_{D}^{0}\eta -iL_{H}\lambda -\frac{i}{2}%
L_{K}^{0}(\eta {\eta ^{\dagger }})\eta \right] {s}(\theta ),  \label{7} \\
L^{S} &=&\left[ d\lambda +\frac{1}{2}d\eta (\lambda {\lambda ^{\dagger }}%
)-\lambda (\lambda d{\eta ^{\dagger }})-L_{H}\frac{i}{2}(\lambda {\lambda
^{\dagger }})\lambda -L_{D}^{0}\frac{1}{2}\left( \lambda +(\lambda {\eta
^{\dagger }})\lambda -\frac{1}{2}\eta (\lambda {\lambda ^{\dagger }})\right)
\right.  \notag \\
&&\left. +L_{K}^{0}\{-i\eta +(\eta \sigma _{a}{\eta ^{\dagger }})\frac{i}{2}%
\lambda \sigma _{a}-\frac{i}{2}(\eta {\eta ^{\dagger }})(\lambda {\eta
^{\dagger }})\lambda -\frac{i}{4}\eta (\eta {\eta ^{\dagger }})(\lambda {%
\lambda ^{\dagger }})\}\right] s(\theta ).  \label{11}
\end{eqnarray}%
$L^{Q^{\dagger }}$ and $L^{S^{\dagger }}$ are conjugate to $L^{Q}$ and $%
L^{S} $ respectively. $L_{H,D,K}^{0}$ are the Maurer-Cartan forms associated
to the SO(1,2),
\begin{equation}
L_{H}^{0}=-e^{-z}dt,\qquad L_{D}^{0}=2e^{-z}\omega dt+dz,\qquad
L_{K}^{0}=-e^{-z}\omega ^{2}dt-\omega dz+d\omega  \label{SOCOSET}
\end{equation}%
while those of the SU(2) are
\begin{equation}
L_{a}^{0}=d\theta ^{b}\;{\mathbf{L}_{ba}},\qquad {\mathbf{L}}_{ba}=\left(
\begin{array}{ccc}
\cos \theta ^{2}\cos \theta ^{3} & \cos \theta ^{2}\sin \theta ^{3} & -\sin
\theta ^{2} \\
-\sin \theta ^{3} & \cos \theta ^{3} & 0 \\
0 & 0 & 1%
\end{array}%
\right) .  \label{sucoset}
\end{equation}%
$s(\theta )$ and $\mathcal{S}_{ab}(\theta )$ are spinor and adjoint
representations of the SU(2) rotation $e^{i\theta ^{a}J_{a}}$ given in (\ref%
{adrep}) and (\ref{spinrep}) in the appendix A respectively.

\textbf{SU(2) matrices}

The group element $g_J$ in the SU(2) sector (\ref{gpara}) is
\begin{equation}
{g_J=e}^{i{\theta ^{1}}J_{1}}{e}^{i{\theta ^{2}}J_{2}} {e}^{i{\theta ^{3}}%
J_{3}}  \label{GJ}
\end{equation}%
$\mathcal{S}_{ab}$ is the adjoint representation of the $g_J$
\begin{equation}
\mathcal{S}_{ab}=\left(
\begin{array}{ccc}
\cos \theta ^{2}\cos \theta ^{3} & \cos \theta ^{2}\sin \theta ^{3} & -\sin
\theta ^{2} \\
\sin \theta ^{1}\sin \theta ^{2}\cos \theta ^{3}-\cos \theta ^{1}\sin \theta
^{3} & \sin \theta ^{1}\sin \theta ^{2}\sin \theta ^{3}+\cos \theta ^{1}\cos
\theta ^{3} & \sin \theta ^{1}\cos \theta ^{2} \\
\cos \theta ^{1}\sin \theta ^{2}\cos \theta ^{3}+\sin \theta ^{1}\sin \theta
^{3} & \cos \theta ^{1}\sin \theta ^{2}\sin \theta ^{3}-\sin \theta ^{1}\cos
\theta ^{3} & \cos \theta ^{1}\cos \theta ^{2}%
\end{array}%
\right)  \label{adrep}
\end{equation}%
while the spinorial representation is:
\begin{equation}
{s=e}^{i\frac{\theta ^{1}}{2}\sigma _{1}}{e}^{i\frac{\theta ^{2}}{2}\sigma
_{2}}{e}^{i\frac{\theta ^{3}}{2}\sigma _{3}}.  \label{spinrep}
\end{equation}%
It holds
\begin{equation}
s^\dagger\sigma_as=\mathcal{S}_{ab}\sigma_b,\qquad (\mathcal{S}^T)_{ad}d%
\mathcal{S}_{db}=\epsilon_{abc}L_c^0,\qquad s^\dagger d s=\frac{i}{2}%
L_c^0\sigma_c.  \label{u2con2}
\end{equation}

The SU(2) left invariant one forms (\ref{sucoset}) are
\begin{equation}
L_{J_{a}}^{0}=d\theta ^{b}\;{\mathbf{L}_{ba}},\qquad {\mathbf{L}}=\left(
\begin{array}{ccc}
\cos \theta ^{2}\cos \theta ^{3} & \cos \theta ^{2}\sin \theta ^{3} & -\sin
\theta ^{2} \\
-\sin \theta ^{3} & \cos \theta ^{3} & 0 \\
0 & 0 & 1%
\end{array}%
\right)  \label{LL}
\end{equation}%
The right invariant one forms defined by $%
-idg_{J}g_{J}^{-1}=J_{a}R_{J_{a}}^{0}$ are
\begin{equation}
R_{J_{a}}^{0}=d\theta ^{b}\;\mathbf{R}_{ba},\qquad \mathbf{R}=\left(
\begin{array}{ccc}
1 & 0 & 0 \\
0 & \cos \theta ^{1} & -\sin \theta ^{1} \\
-\sin \theta ^{2} & -\sin \theta ^{1}\cos \theta ^{2} & \cos \theta ^{1}\cos
\theta ^{2}%
\end{array}%
\right) .
\end{equation}%
The matrix $\mathbf{R}_{ab}^{-1}$
%appearing SU(2) transformations  \bref{SU2Rot}
is the inverse of $\mathbf{R}$;

\begin{equation}
\mathbf{R}^{-1}=\left(
\begin{array}{ccc}
1 & 0 & 0 \\
\sin \theta ^{1}\tan \theta ^{2} & \cos \theta ^{1} & \sin \theta ^{1}/\cos
\theta ^{2} \\
\cos \theta ^{1}\tan \theta ^{2} & -\sin \theta ^{1} & \cos \theta ^{1}/\cos
\theta ^{2}%
\end{array}%
\right) =\mathcal{S}\;\mathbf{L}^{-1}.  \label{bRinv}
\end{equation}

When the lagrangian was constructed in (\ref{suinv}), the following
shorthands are used
\begin{eqnarray}
N_{H} &=&b_{H}+b_{K}\frac{1}{4}(\lambda {\lambda ^{\dagger }})^{2}-(\lambda
\sigma _{a}{\lambda ^{\dagger }})\mathcal{S}_{ab}b_{b}  \label{NH} \\
N_{D} &=&b_{D}\{1+\frac{1}{2}(\lambda {\eta ^{\dagger }}+\eta {\lambda
^{\dagger }})\}-b_{K}\frac{i}{4}(\lambda {\eta ^{\dagger }}-\eta {\lambda
^{\dagger }})(\lambda {\lambda ^{\dagger }})+\frac{i}{2}\left( \lambda
\sigma _{a}{\eta ^{\dagger }}-\eta \sigma _{a}{\lambda ^{\dagger }}\right)
\mathcal{S}_{ab}b_{b}  \label{ND} \\
N_{K} &=&b_{K}\{1+(\lambda {\eta ^{\dagger }}+\eta {\lambda ^{\dagger }})-%
\frac{1}{2}(\eta \sigma _{a}{\eta ^{\dagger }})(\lambda \sigma _{a}{\lambda
^{\dagger }})+\frac{1}{4}(\eta {\eta ^{\dagger }})(\lambda {\lambda
^{\dagger }})(\lambda {\eta ^{\dagger }}+\eta {\lambda ^{\dagger }})+\frac{1%
}{16}(\eta {\eta ^{\dagger }})^{2}(\lambda {\lambda ^{\dagger }})^{2}\}
\notag \\
&&+b_{H}\frac{1}{4}(\eta {\eta ^{\dagger }})^{2}+b_{D}\frac{i}{2}(\eta {\eta
^{\dagger }})(\lambda {\eta ^{\dagger }}-\eta {\lambda ^{\dagger }})  \notag
\\
&&-\{(\eta \sigma _{a}{\eta ^{\dagger }})+\frac{1}{2}(\eta {\eta ^{\dagger }}%
)\left( \lambda \sigma _{a}{\eta ^{\dagger }}+\eta \sigma _{a}{\lambda
^{\dagger }}\right) +\frac{1}{4}(\eta {\eta ^{\dagger }})^{2}(\lambda \sigma
_{a}{\lambda ^{\dagger }})\}\mathcal{S}_{ab}b_{b}  \label{NK}
\end{eqnarray}%
\begin{eqnarray}
N_{rest} &=&b_{H}\{-\frac{i}{2}(\eta d{\eta ^{\dagger }}-d\eta \eta
^{\dagger })\}+b_{D}\{(\lambda d{\eta ^{\dagger }}+d\eta {\lambda ^{\dagger }%
})\}  \notag \\
&+&b_{K}\{-\frac{i}{2}(\lambda d\lambda ^{\dagger }-d\lambda \lambda
^{\dagger })-\frac{i}{2}(\lambda d\eta ^{\dagger }-d\eta \lambda ^{\dagger
})(\lambda \lambda ^{\dagger })-\frac{i}{8}(\eta d\eta ^{\dagger }-d\eta
\eta ^{\dagger })(\lambda \lambda ^{\dagger })^{2}\}  \notag \\
&+&\{i(\lambda \sigma _{a}d{\eta ^{\dagger }}-d\eta \sigma _{a}{\lambda
^{\dagger }})+\frac{i}{2}(\eta d{\eta ^{\dagger }}-d\eta {\eta ^{\dagger }}%
)(\lambda \sigma _{a}{\lambda ^{\dagger }})\}\mathcal{S}_{ab}b_{b}.
\label{Nres}
\end{eqnarray}

\textbf{MC forms in the AdS basis}

The bosonic part of the MC forms in the AdS basis are given as
\begin{eqnarray}
L_0^{P_0}&=&dx^0\cosh\frac{x^1}{R},\qquad L_0^{P_1}=dx^1,\qquad L_0^{M_{01}}=%
\frac{dx^0}{R}\sinh\frac{x^1}{R},  \label{adsL00} \\
L_0^{J_1}&=&d\theta^1\cos\theta^2,\qquad L_0^{J_2}=d\theta^2,\qquad
L_0^{J_3}=-d\theta^1\sin\theta^2.  \label{adsL0}
\end{eqnarray}
The $L_1^{P_0},L_1^{P_1},L_1^{M_{01}},L_1^{J_b}$ in (\ref{ABC}) are
including the fermionic contributions and are obtained as
%from  \bref{1}-\bref{3},
\begin{eqnarray}
L_1^{P_0}&=&R(L_1^H+L_1^K),\qquad L_1^{P_1}=RL_1^D,\qquad
L_1^{M_{01}}=L_1^K-L_1^H,\qquad L_1^{J_{b}}.
\end{eqnarray}
Here $L_1^H,L_1^K,L_1^D,L_1^{J_b}$ are obtained from $L^H,L^K,L^D,L^{J_b}$
in the conformal basis (\ref{1})-(\ref{3}) in which $%
L^0_H,L^0_K,L^0_D,L^0_{J_b}$ are replaced by
\begin{eqnarray}
L^0_H\to\frac{L_0^{P_0}}{2R}-\frac{L_0^{M_{01}}}{2},\qquad L^0_K\to\frac{%
L_0^{P_0}}{2R}+\frac{L_0^{M_{01}}}{2},\qquad L^0_D\to\frac{L_0^{P_1}}{R}%
,\qquad L^0_{J_b}\to L_0^{J_b},
\end{eqnarray}
where the bosonic MC one forms in the AdS base are given in (\ref{adsL00})
and (\ref{adsL0}).

\section{Appendix: PSU(1,1$|$2) transformations}

The bosonic transformations of $PSU(1,1|2)$ are given in (\ref{D})-(\ref{TD}%
). Supersymmetric and superconformal transformations of the goldstone fields
can be calculated in the same way, obtaining however, complicated
expressions. It is convenient to give them here for further references.

\begin{itemize}
\item Ordinary supersymmetry:%
\begin{align}
\delta _{Q}t& =\frac{i}{2}e^{z/2}\left[ -\eta ^{\dagger }+i\frac{1}{2}\omega
\left( \eta \eta ^{\dagger }\right) \left( \eta ^{\dagger }-\frac{\lambda
^{\dagger }}{2\Delta }\left( \eta \eta ^{\dagger }\right) \right) \right]
\epsilon _{Q}  \notag \\
\delta _{Q}z& =\frac{1}{2}\omega ^{2}e^{-z/2}\left( \eta \eta ^{\dagger
}\right) \left( \eta ^{\dagger }-\frac{\lambda ^{\dagger }}{2\Delta }\left(
\eta \eta ^{\dagger }\right) \right) \epsilon _{Q}  \notag \\
\delta _{Q}\omega & =-\frac{i}{2}e^{-z/2}\omega ^{2}\left[ \eta ^{\dagger }+i%
\frac{1}{2}\omega \left( \eta \eta ^{\dagger }\right) \left( \eta ^{\dagger
}-\frac{\lambda ^{\dagger }}{2\Delta }\left( \eta \eta ^{\dagger }\right)
\right) \right] \epsilon _{Q}-\frac{\omega \lambda ^{\dagger }}{2\Delta }%
e^{-z/2}\epsilon _{Q} \\
\delta _{Q_{i}}\eta _{k}& =e^{-z/2}\left( \delta _{ik}+\frac{i}{2}\omega
\left( \eta \eta ^{\dagger }\right) \left( -\delta _{ik}+\frac{\lambda
_{i}^{\dagger }\eta _{k}}{\Delta }\right) \right) \epsilon _{Q_{i}}  \notag
\\
\delta _{Q_{i}}\eta _{k}^{\dagger }& =i\omega e^{-z/2}\left( \eta ^{\dagger
}-\frac{\lambda _{i}^{\dagger }}{\Delta }\left( \eta \eta ^{\dagger }\right)
\right) \eta _{k}^{\dagger }\epsilon _{Q_{i}}  \notag \\
\delta _{Q_{i}}\lambda _{k}& =i\omega e^{-z/2}\left( \delta _{ik}-\frac{%
\lambda _{i}^{\dagger }\eta _{k}}{2\Delta }\right) \epsilon _{Q_{i}}  \notag
\\
\delta _{Q_{i}}\lambda _{k}^{\dagger }& =i\omega e^{-z/2}\frac{\lambda
_{i}^{\dagger }\eta _{k}^{\dagger }}{2\Delta }\epsilon _{Q_{i}}
\label{susy1} \\
\delta _{Q}\theta ^{b}& =-i\omega e^{-z/2}\left( i\sigma _{a}\eta ^{\dagger
}-\frac{\lambda ^{\dagger }}{2\Delta }\left( i\eta \sigma _{a}\eta ^{\dagger
}\right) \right)\mathcal{P}^{ab}\epsilon _{Q_{i}}
\end{align}
\end{itemize}

Where
\begin{equation}
\bigskip \Delta =1+\frac{\eta \lambda ^{\dagger }+\lambda \eta ^{\dagger }}{2}.
\end{equation}
The transformations under $Q^\dagger$ are obtained by taking conjugations.
For example from (\ref{susy1})
\begin{eqnarray}
\delta _{Q_{i}^{\dagger }}\eta _{k}^{\dagger }& =e^{-z/2}\left( \delta
_{ik}+ \frac{i}{2}\omega \left( \eta \eta ^{\dagger }\right) \left(\delta
_{ik}+\frac{\lambda _{i}\eta _{k}^{\dagger }}{\Delta }\right) \right)
\epsilon _{Q_{i}^{\dagger }} .
\end{eqnarray}

\begin{itemize}
\item \bigskip Superconformal transformations%
\begin{eqnarray}
\delta _{S}t &=&\frac{i}{2}e^{z/2}\left[ it\eta ^{\dagger }+\frac{1}{2}%
\left( e^{z}+t\omega \right) \left( \eta \eta ^{\dagger }\right) \left( \eta
^{\dagger }-\frac{\lambda ^{\dagger }}{2\Delta }\left( \eta \eta ^{\dagger
}\right) \right) \right] \epsilon _{S}  \notag \\
\delta _{S}z &=&-\frac{i}{2}\omega e^{-z/2}\left( e^{z}+t\omega \right)
\left( \eta \eta ^{\dagger }\right) \left( \eta ^{\dagger }-\frac{\lambda
^{\dagger }}{2\Delta }\left( \eta \eta ^{\dagger }\right) \right) \epsilon
_{S}-e^{z/2}\eta ^{\dagger }\epsilon _{S}  \notag \\
\delta _{S}\omega &=&\frac{i}{2}e^{-z/2}\omega ^{2}\left[ it\eta ^{\dagger }-%
\frac{1}{2}\left( e^{z}+t\omega \right) \left( \eta \eta ^{\dagger }\right)
\left( \eta ^{\dagger }-\frac{\lambda ^{\dagger }}{2\Delta }\left( \eta \eta
^{\dagger }\right) \right) \right] \epsilon _{S}-e^{z/2}\eta ^{\dagger
}\omega \epsilon _{S}  \notag \\
&&+i\frac{\lambda ^{\dagger }}{2\Delta }\left( e^{z/2}+t\omega
e^{-z/2}\right) \epsilon _{S} \\
\delta _{S_{i}}\eta _{k} &=&\left( -ite^{-z/2}\delta _{ik}+\frac{1}{2}\left(
e^{z/2}+t\omega e^{-z/2}\right) \left( \eta \eta ^{\dagger }\right) \left(
-\delta _{ik}+\frac{\lambda _{i}^{\dagger }\eta _{k}}{\Delta }\right)
\right) \epsilon _{S_{i}}  \label{suco1} \\
\delta _{S_{i}}\eta _{k}^{\dagger } &=&\left( e^{z/2}+t\omega
e^{-z/2}\right) \left( \eta \eta ^{\dagger }\right) \left( \eta
_{i}^{\dagger }-\frac{\lambda _{i}^{\dagger }}{\Delta }\left( \eta \eta
^{\dagger }\right) \right) \eta _{k}^{\dagger }\epsilon _{S_{i}}  \notag \\
\delta _{S_{i}}\lambda _{k} &=&\left( e^{z/2}+t\omega e^{-z/2}\right) \left(
\delta _{ik}-\frac{\lambda _{i}^{\dagger }\eta _{k}}{2\Delta }\right)
\epsilon _{S_{i}}  \notag \\
\delta _{S_{i}}\lambda _{k}^{\dagger } &=&\left( e^{z/2}+t\omega
e^{-z/2}\right) \frac{\lambda _{i}^{\dagger }\eta _{k}^{\dagger }}{2\Delta }%
\epsilon _{S_{i}}  \notag \\
\delta _{S}\theta ^{b} &=&-\left( e^{z/2}+t\omega e^{-z/2}\right) \left(
i\sigma _{a}\eta ^{\dagger }-\frac{\lambda ^{+}}{2\Delta }\left( i\eta
\sigma _{a}\eta ^{\dagger }\right) \right)\mathcal{P}^{ab}\epsilon _{S}
\notag \\
\end{eqnarray}
\end{itemize}

%%%%%%%%%%%%%%%%%%%%%%%%%%%%%%%%%%%%%%%%%%%%%%%%%%%%%%%%%%

\section{Appendix: Diffeomorphism in terms of the gauge symmetries}

It is shown here that the diffeomorphism of the action (\ref{SUlagr}) is
equivalent to a suitable combination of the $T$-gauge (\ref{Htrans}), $U(1)$
(\ref{U1trans}) and kappa (\ref{kaptr1}) transformations for the BPS case $%
b_{H}b_{K}=b_{D}^{2}+b_{a}b_{a}$, (\ref{kappacond}).

\subsection{Trivial Symmetry}

The Euler derivatives $(\mathcal{L})_{M}$ are defined as%
\begin{equation}
\delta \mathcal{L}=(\mathcal{L})_{M}\delta Z^{M}+\mathrm{surface\;term},
\label{Evari}
\end{equation}%
any action is invariant under a transformation of the form
\begin{equation}
\delta Z^{M}=(\mathcal{L})_{N}A^{NM},\qquad A^{MN}=-(-)^{MN}A^{NM},
\end{equation}%
that is, $A^{MN}$ is graded anti-symmetric. $(-)^{MN}=-1$ when both $M$ and $%
N$ are odd indices and $(-)^{MN}=+1$ otherwise. It is a trivial symmetry and
does not lead to a Noether charge. Now the Lagrangian is (\ref{SUlagr}).
\begin{equation}
\mathcal{L}=b_{A}L^{A}.
\end{equation}%
In this appendix the pullback on $L^{A}$ is tacitly understood. The Euler
derivative $(\mathcal{L})_{M}$ is
\begin{equation}
(\mathcal{L})_{M}=b_{A}f_{BC}^{A}(\dot{Z}^{N}{L_{N}}^{C})({L_{M}}%
^{B})(-1)^{M(M+B)}.
\end{equation}
In the present formulation we use all group coordinates $Z^{M}$ the ${%
L_{M}^{\prime }}^{B}\equiv ({L_{M}}^{B})(-1)^{M(M+B)}$ has the inverse ${%
L_{B}^{\prime }}^{M}$. It is convenient to define
\begin{equation}
\lbrack \mathcal{L}]_{B}=(\mathcal{L})_{M}{L_{B}^{\prime }}%
^{M}=b_{A}f_{BC}^{A}(\dot{Z}^{N}{L_{N}}^{C}).  \label{bED}
\end{equation}%
Using it (\ref{Evari}) is written as
\begin{equation}
\delta \mathcal{L}=[\mathcal{L}]_{A}[\delta Z^{A}]+\mathrm{surface\;term}.
\label{Evari2}
\end{equation}%
Then a transformation is trivial if $[\delta Z^{A}]$ is written as a
(graded) antisymmetric combination of the equations of motion (\ref{bED}),
\begin{equation}
\lbrack \delta Z^{A}]=[\mathcal{L}]_{B}\tilde{A}^{BA},\qquad \tilde{A}%
^{AB}=-(-)^{AB}\tilde{A}^{BA}.
\end{equation}%
\vskip3mm

\subsection{Geometrical diffeomorphism}

For the geometrical diffeomorphism
\begin{equation}
\delta_{diff} Z^M=\varepsilon \dot Z^M,\quad \to \quad \left[\delta_{diff}
Z^A\right]=\varepsilon \dot Z^M{\ L_M}^A=\varepsilon L^A
\end{equation}
We will show the geometrical diffeomorphism is not independent of the gauge
transformations but equivalent to a combination of the gauge
transformations. More precisely they differ by a trivial transformation
discussed above.

The gauge transformations of (3.3-6) is
\begin{eqnarray}
\left[ \delta _{gauge}t\right] &=&\epsilon (\tau ),\quad \left[ \delta
_{gauge}z\right] =-2\frac{b_{D}}{b_{K}}\epsilon (\tau ),\quad \left[ \delta
_{gauge}w\right] =\frac{b_{H}}{b_{K}}\epsilon (\tau ),\quad  \notag \\
\left[ \delta _{gauge}\theta ^{a}\right] &=&{b_{J_{a}}}\alpha (\tau ),
\notag \\
\left[ \delta _{gauge}\eta \right] &=&\kappa _{\eta }(\tau )s(\theta
),\qquad \left[ \delta _{gauge}\lambda \right] =\kappa _{\eta }(\tau
)s(\theta )(\frac{ib_{D}}{b_{K}}+\frac{b_{a}\sigma _{a}}{b_{K}}).
\label{kaptr2}
\end{eqnarray}%
Let $\Delta $ is difference of ``$\delta _{diff}$" and ``$\delta _{gauge}$",
\begin{eqnarray}
\left[ \Delta t\right] &=&=\varepsilon L^{H}-\epsilon (\tau ),\quad \left[
\Delta z\right] =\varepsilon L^{D}+2\frac{b_{D}}{b_{K}}\epsilon (\tau ),\quad
\notag \\
\left[ \Delta w\right] &=&\varepsilon L^{K}-\frac{b_{H}}{b_{K}}\epsilon
(\tau ),\quad \left[ \Delta \theta ^{a}\right] =\varepsilon L^{J_{a}}-{%
b_{J_{a}}}\alpha (\tau ),  \notag \\
\left[ \Delta \eta \right] &=&\varepsilon L^{Q}-\kappa _{\eta }(\tau
)s(\theta ),\qquad \left[ \Delta \lambda \right] =\varepsilon L^{S}-\kappa
_{\eta }(\tau )s(\theta )(\frac{ib_{D}}{b_{K}}+\frac{b_{a}\sigma _{a}}{b_{K}}%
),\qquad  \label{kaptr3}
\end{eqnarray}%
We choose the gauge parameter functions $\epsilon ,\alpha ,\kappa $ as
\begin{equation}
\epsilon (\tau )=\varepsilon L^{H},\qquad \alpha (\tau )=\varepsilon \frac{%
(b_{b}L^{b})^{\ast }}{b_{c}^{2}},\qquad \kappa _{\eta }(\tau )s(\theta
)=\varepsilon L^{Q}  \label{kaptr312}
\end{equation}%
so that, using Euler derivatives in (\ref{bED}),
\begin{eqnarray}
\left[ \Delta t\right] &=&0,\quad  \notag \\
\left[ \Delta z\right] &=&\frac{\varepsilon }{b_{K}}\left( b_{K}L^{D}+2{b_{D}%
}{L^{H}}\right) =-\frac{\varepsilon }{b_{K}}[\mathcal{L}]_{K},\quad  \notag
\\
\left[ \Delta w\right] &=&\frac{\varepsilon }{b_{K}}\left( b_{K}L^{K}-{b_{H}}%
{L^{H}}\right) =\frac{\varepsilon }{b_{K}}[\mathcal{L}]_{D},\quad  \notag \\
\left[ \Delta \theta ^{a}\right] &=&\frac{\varepsilon }{{b_{c}^{2}}}\epsilon
_{abc}b_{c}(\epsilon _{bde}b_{d}L^{e})=-\frac{\varepsilon }{{b_{c}^{2}}}%
\epsilon _{abc}b_{c}\;[\mathcal{L}]_{b},  \notag \\
\left[ \Delta \eta \right] &=&0,\qquad  \notag \\
\left[ \Delta \lambda \right] &=&\frac{\varepsilon }{b_{K}}\left(
b_{K}L^{S}-L^{Q}({ib_{D}}+{b_{a}\sigma _{a}})\right) =-i\frac{\varepsilon }{%
b_{K}}\;[\mathcal{L}]_{S}.  \label{kaptr5}
\end{eqnarray}%
We also have for the conjugate coordinates
\begin{equation}
\left[ \Delta \eta ^{\dagger }\right] =0,\qquad \left[ \Delta \lambda
^{\dagger }\right] =i\frac{\varepsilon }{b_{K}}\;[\mathcal{L}]_{S^{\dagger
}}.
\end{equation}%
From (\ref{gpara}) remembering that the coordinate for $S$ is $\lambda
^{\dagger }$ while that of $S^{\dagger }$ is $-\lambda$ they are written in
the matrix form
\begin{equation}
\begin{pmatrix}
\lbrack \Delta t]\cr [\Delta z ]\cr [\Delta w ]\cr [\Delta \theta_a ]\cr
[\Delta \eta^\dagger ]\cr-[\Delta \eta ]\cr [\Delta \lambda^\dagger ]\cr%
-[\Delta \lambda ]%
\end{pmatrix}%
^{T}=%
\begin{pmatrix}
\left[ \mathcal{L}\right] _{H}\cr\left[ \mathcal{L}\right] _{D}\cr\left[
\mathcal{L}\right] _{K}\cr\left[ \mathcal{L}\right] _{J_{b}}\cr\left[
\mathcal{L}\right] _{Q}\cr\left[ \mathcal{L}\right] _{Q^{\dagger }}\cr\left[
\mathcal{L}\right] _{S}\cr\left[ \mathcal{L}\right] _{S^{\dagger }}%
\end{pmatrix}%
^{T}%
\begin{pmatrix}
. & . & . & . & . & . & . & .\cr. & . & \frac{\varepsilon }{b_{K}} & . & . &
. & . & .\cr. & -\frac{\varepsilon }{b_{K}} & . & . & . & . & . & .\cr. & .
& . & -\frac{\varepsilon }{{b_{d}^{2}}}\epsilon _{abc}b_{c} & . & . & . & .%
\cr. & . & . & . & . & . & . & .\cr. & . & . & . & . & . & . & .\cr. & . & .
& . & . & . & . & i\frac{\varepsilon }{b_{K}}\cr. & . & . & . & . & . & i%
\frac{\varepsilon }{b_{K}} & .%
\end{pmatrix}%
.
\end{equation}%
The matrix appearing here is graded anti-symmetric and the transformation $%
\Delta $ is shown to be trivial.

In the non-BPS case, $b_{H}b_{K}- b_{D}^{2}\neq b_{J_a}^2$, there is no
kappa symmetry and $\kappa_\eta$ is taken to be zero in (\ref{kaptr3}) and (%
\ref{kaptr312}). In this case
\begin{eqnarray}
\left[\Delta \eta \right]&=&\varepsilon L^Q= -i\varepsilon\frac{\left[%
\mathcal{L} \right]_Qb_K-\left[\mathcal{L} \right]_S ({ib_{D}}-{b_{a}\sigma
_{a}})}{b_{H}b_{K}- b_{D}^{2}- b_{J_a}^2} ,\qquad  \notag \\
\left[\Delta \lambda \right]&=&\varepsilon L^S= -i\varepsilon\frac{\left[%
\mathcal{L} \right]_Sb_H+\left[\mathcal{L} \right]_S ({ib_{D}}+{b_{a}\sigma
_{a}})}{b_{H}b_{K}- b_{D}^{2}- b_{J_a}^2}.  \label{kaptrnBPS}
\end{eqnarray}
They are also graded anti-symmetric combinations of the equations of motion
and the difference of the diffeomorphism and the $H$ and $U(1)$
transformations is a trivial transformation.

%%%%%%%%%%%%%%%%%%%%%%%%%%%%%%%%%%%%%%%%%%%%%%%%%%%%%%%%%%

\section{Appendix: Conformal mechanics invariant under $OSP(2|2)$}

In this appendix we explicitly derive the kappa transformation of the $%
OSP(2|2)$ case in an arbitrary configuration. Furthermore kappa invariant
and quasi invariant\textbf{\ } variables are constructed and the lagrangian
is written in terms of them. To show the relation with the former case a
dictionary is given.

The OSP(2$|$2) is a subalgebra of $SU(\left. 1,1\right\vert 2)$ whose
generators are
\begin{equation}
H,\quad K,\quad D,\quad B=-2J_2
\end{equation}
and
\begin{equation}
\mathbf{Q}_i=\frac{1}{\sqrt{2}}(Q^i+{Q^\dagger}_i),\qquad \mathbf{S}_i=\frac{%
i}{\sqrt{2}}(S^i-{S^\dagger}_i).
\end{equation}
They satisfy $OSP(2|2)$ algebra:
\begin{eqnarray}
\left[ H,D\right] &=&iH\qquad \left[ K,D\right] =-iK\qquad \left[ H,K\right]
=2iD \\
\left[ \mathbf{Q}_{i},\mathbf{Q}_{j}\right] _{+} &=&\delta _{ij}H\qquad %
\left[ \mathbf{S}_{i},\mathbf{S}_{j}\right] _{+}=\delta _{ij}K\qquad \left[
\mathbf{Q}_{i},\mathbf{S}_{j}\right] _{+}=\delta _{ij}D+\frac{1}{2}%
\varepsilon _{ij}B \\
\left[ D,\mathbf{Q}_{i}\right] &=&-\frac{i}{2}\mathbf{Q}_{i}\qquad \left[ D,%
\mathbf{S}_{i}\right] =\frac{i}{2}\mathbf{S}_{i}\qquad \left[ K,\mathbf{Q}%
_{i}\right] =-i\mathbf{S}_{i} \\
\left[ H,\mathbf{S}_{i}\right] &=&i\mathbf{Q}_{i}\qquad \left[ B,\mathbf{Q}%
_{i}\right] =-i\varepsilon _{ij}\mathbf{Q}_{j}\qquad \left[ B,\mathbf{S}_{i}%
\right] =-i\varepsilon _{ij}\mathbf{S}_{j}  \label{osp22algebra}
\end{eqnarray}
The group element is parametrized as
\begin{equation}
g=e^{-itH}e^{izD}e^{i\omega K}e^{i\tilde{\eta} Q}e^{i\tilde{\lambda}
S}e^{i\alpha B}
\end{equation}
All formulas of OSP(2$|$2) must be given from those of the SU(1,1$|$2) by
the following replacements
\begin{eqnarray}
\eta^i&\to& -\frac{\tilde\eta^i}{\sqrt{2}},\qquad \eta^\dagger_i\to \frac{%
\tilde\eta^i}{\sqrt{2}},\qquad \tilde\eta=\frac{1}{\sqrt{2}}%
(\eta^\dagger_i-\eta^i)  \notag \\
\lambda^i&\to& \frac{i}{\sqrt{2}}\tilde\lambda^i,\qquad \lambda^\dagger_i\to
\frac{i}{\sqrt{2}}\tilde\lambda^i,\qquad \tilde\lambda=\frac{-i}{\sqrt{2}}%
(\lambda^\dagger_i+\lambda^i)  \notag \\
\quad \theta^2&=&-2\alpha,\quad and \quad \theta^1=\theta^3=0.
\end{eqnarray}

The components of the left invariant Maurer-Cartan form are:
\begin{eqnarray}
L^{H} &=&{-e^{-z}dt}+\frac{i}{2}(\eta d\eta )  \notag \\
L^{K} &=&d\omega \left( 1+i(\lambda \eta )+\frac{1}{8}(\lambda \epsilon
\lambda )(\eta \epsilon \eta )\right) -\omega dz\left( 1+i(\lambda \eta )+%
\frac{1}{8}(\lambda \epsilon \lambda )(\eta \epsilon \eta )\right)  \notag \\
&&-\omega ^{2}{e^{-z}dt}\left( 1+i(\lambda \eta )+\frac{1}{8}(\lambda
\epsilon \lambda )(\eta \epsilon \eta )\right) +\frac{i}{2}(\lambda d\lambda
)  \notag \\
L^{D} &=&dz\left( 1+\frac{i}{2}(\lambda \eta )\right) +2\omega {e^{-z}dt}%
\left( 1+\frac{i}{2}(\lambda \eta )\right) +{i}(\lambda d\eta )  \notag \\
L^{B} &=&d\alpha +\frac{i}{4}\;d\omega \left( \eta \epsilon \eta \right) -%
\frac{i}{4}\;dz\left( (\lambda \epsilon \eta )+\omega (\eta \epsilon \eta
)\right)  \notag \\
&&-{e^{-z}dt}\frac{i}{4}\left( (\lambda \epsilon \lambda )+2\omega (\lambda
\epsilon \eta )+\omega ^{2}(\eta \epsilon \eta )\right)  \notag \\
&&-\frac{i}{2}(\lambda \epsilon d\eta )-\frac{1}{8}(\lambda \epsilon \lambda
)(\eta d\eta )  \notag \\
L^{Q} &=&(\cos \alpha +\epsilon \;\sin \alpha )\left[ d\eta +\frac{1}{2}\eta
\;dz+{e^{-z}dt}(\lambda +\omega \eta )-\frac{i}{2}\;\lambda (\eta d\eta )%
\right] ,  \notag \\
L^{S} &=&(\cos \alpha +\epsilon \;\sin \alpha )\left[ d\lambda +(\eta -\frac{%
i}{4}\epsilon \lambda (\eta \epsilon \eta ))\;d\omega -dz\left( \frac{1}{2}%
\lambda +\eta \omega +\frac{i}{8}\epsilon \eta (\lambda \epsilon \lambda )-%
\frac{i}{4}\epsilon \lambda \omega (\eta \epsilon \eta )\right) \right.
\notag \\
&&\left. -{e^{-z}dt}\left( \omega \lambda +\omega ^{2}\eta +\frac{i}{4}%
\omega \epsilon \eta (\lambda \epsilon \lambda )-\frac{i}{4}\omega
^{2}\epsilon \lambda (\eta \epsilon \eta )\right) -\frac{i}{4}\;\epsilon
d\eta (\lambda \epsilon \lambda )\right] ,
\end{eqnarray}
where $\epsilon=i\sigma_2$.

Now the action is:
\begin{eqnarray}
\mathcal{S} &=&\int \mathcal{L}d\tau =\int \left(
b_{H}L^{H}+b_{K}L^{K}+b_{D}L^{D}+b_{B}L^{B}\right) ^{\ast }  \label{OSP22L}
\end{eqnarray}

Under $\kappa $ variations satisfying
\begin{equation}
\left[ \delta t\right] =\left[ \delta z\right] =\left[ \delta \alpha \right]
=\left[ \delta \omega \right] =0  \label{ospgold}
\end{equation}%
the LI one forms transform as:
\begin{equation}
\delta L^{H}=-iL^{Q}\left[ \delta \eta \right] \qquad \delta L^{K}=-iL^{S}%
\left[ \delta \lambda \right] \qquad \delta L^{D}=-i\left( L^{Q}\left[
\delta \lambda \right] +L^{S}\left[ \delta \eta \right] \right)
\end{equation}%
\begin{equation}
\delta L^{B}=-\frac{i}{2}\left( L^{Q}\epsilon \left[ \delta \lambda \right]
-L^{S}\epsilon \left[ \delta \eta \right] \right).
\end{equation}%
The condition for the lagrangian $\left( \ref{OSP22L}\right) $ to be kappa
invariant is given by:
\begin{equation}
\left[ \delta \eta \right] =-\frac{1}{b_{H}}\left( b_{D}+\frac{1}{2}%
b_{B}\epsilon \right) \left[ \delta \lambda \right] \qquad \left[ \delta
\lambda \right] =-\frac{1}{b_{K}}\left( b_{D}-\frac{1}{2}b_{B}\epsilon
\right) \left[ \delta \eta \right]  \label{kappa ferm cond}
\end{equation}%
which in turn implies:
\begin{equation}
b_{K}b_{H}=b_{D}^{2}+\frac{1}{4}b_{B}^{2}.  \label{k}
\end{equation}%
When this is verified it is kappa symmetric else it describes non-BPS
paricle.

The kappa transformations of the BPS particle are found as follows. Due to
the former condition, (\ref{ospgold}), we can find the explicit form of the
kappa variations for the bosonic fields in term of the fermionic ones:
\begin{eqnarray}
\delta _{\kappa }t &=&\frac{i}{2}e^{z}\eta \delta _{\kappa }\eta  \notag \\
\delta _{\kappa }z &=&-\left( 1-\frac{i}{2}(\lambda \eta )\right) i\lambda
\delta _{\kappa }\eta -\omega i\eta \delta _{\kappa }\eta  \notag \\
\delta _{\kappa }\omega &=&-\left( 1-{i}(\lambda \eta )\right) \frac{i}{2}%
\lambda \delta _{\kappa }\lambda -\left( 1-\frac{i}{2}(\lambda \eta )\right)
i\omega \lambda \delta _{\kappa }\eta -\frac{i}{2}\omega ^{2}\eta \delta
_{\kappa }\eta  \notag \\
\delta _{\kappa }\alpha &=&\frac{i}{2}(\lambda \epsilon \delta _{\kappa
}\eta )-\frac{1}{8}\left( \eta \epsilon \eta \right) (\lambda \delta
_{\kappa }\lambda )+\frac{1}{4}\left( \lambda \epsilon \eta \right) (\lambda
\delta _{\kappa }\eta ).
\end{eqnarray}%
Introducing kappa parameters:
\begin{equation}
\left[ \delta \eta \right] =\left( \cos \alpha +\epsilon \sin \alpha \right)
\kappa _{\eta }\qquad \left[ \delta \lambda \right] =\left( \cos \alpha
+\epsilon \sin \alpha \right) \kappa _{\lambda },
\end{equation}%
we get
\begin{equation}
\delta _{\kappa }\eta =\kappa _{\eta }+\frac{i}{2}\eta \left( \lambda \kappa
_{\eta }\right) \qquad \delta _{\kappa }\lambda =\kappa _{\lambda }+\frac{i}{%
2}\eta \left( \lambda \kappa _{\lambda }\right) .
\end{equation}%
(\ref{kappa ferm cond}) is solved for $\kappa _{\lambda }$ as
\begin{equation}
\kappa _{\lambda }=-\frac{1}{b_{K}}(b_{D}-\frac{1}{2}b_{B}\epsilon )\kappa
_{\eta }.
\end{equation}

We can introduce the kappa invariant variables; fermionic coordinates:
\begin{equation}
\Psi =(\lambda +\frac{1}{b_{K}}(b_{D}-\frac{b_{B}}{2}\epsilon )\eta )+\frac{%
ib_{B}}{4b_{K}}(\lambda \eta )\epsilon \eta .
\end{equation}%
and the bosonic coordinate:
\begin{equation}
q=\sqrt{2}e^{\frac{z}{2}}\left( \frac{N_{K}}{b_{K}}\right) ^{\frac{1}{2}}.
\end{equation}%
Using the kappa condition $\left( \ref{k}\right) $ the lagrangian is
expressed in terms of the kappa invariant variables:%
\begin{equation}
\mathcal{L}=b_{K}\frac{\dot{q}^{2}}{2\dot{t}}-\frac{2\dot{t}}{b_{K}q^{2}}%
\left( \frac{b_{B}^{2}}{4}+\frac{ib_{B}b_{K}}{4}(\Psi \epsilon \Psi )\right)
+b_{K}\frac{i}{2}\Psi \dot{\Psi}+\mathrm{(surface\;term)}.
\end{equation}

\end{document}